\documentclass[rsi,aip,reprint,amsmath,amssymb,floatfix]{revtex4-1}

\usepackage[utf8]{inputenc} % allow utf-8 input
\usepackage[T1]{fontenc}    % use 8-bit T1 fonts
\usepackage[pdfborder={0 0 0.5}]{hyperref}       % hyperlinks
\usepackage{url}            % simple URL typesetting
\usepackage{booktabs}       % professional-quality tables
\usepackage{amsfonts}       % blackboard math symbols
\usepackage{amsmath}
\usepackage{mathtools}
\usepackage{amssymb}
\usepackage{nicefrac}       % compact symbols for 1/2, etc.
\usepackage{lmodern}		% No me compilaba en el portátil sin esto (?)
\usepackage{microtype}      % microtypography
\usepackage{lipsum}		    % Can be removed after putting your text content
\usepackage{natbib}
\usepackage{bibentry}
\usepackage{dcolumn}		% Align table columns on decimal point
\usepackage{physics}
\usepackage{float}
\usepackage{cancel}
\usepackage{graphicx}
\usepackage{xcolor}
\usepackage{csquotes}
\usepackage{placeins}

\graphicspath{{./figs/}}

\newcommand{\orcid}[2]{{#1}{\href{https://orcid.org/#2}{\includegraphics[scale=0.06]{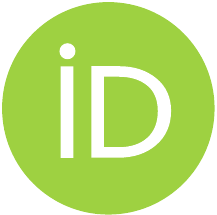}}}}
\newcommand{\ciemat}{Laboratorio Nacional de Fusión - CIEMAT, 28040, Madrid, Spain}
\newcommand{\khipt}{Institute of Plasma Physics of the NSC KIPT, 61108, Kharkiv, Ukraine}

\begin{document}

\title{Exploring the operational limits of poloidal and helical arrays of Mirnov coils in stellarators by means of a synthetic diagnostic}

\author{\orcid{P. Pons-Villalonga}{0000-0003-3225-0592}}
\email[Corresponding author: ]{pedro.pons@ciemat.es}
\affiliation{\ciemat}

\author{\orcid{Á. Cappa}{0000-0002-2250-9209}}
\email[]{alvaro.cappa@ciemat.es}
\affiliation{\ciemat}

\author{\orcid{J. Martínez-Fernández}{0000-0002-5583-8420}}
\affiliation{\ciemat}

\author{O.S. Kozachok}
\affiliation{\khipt}

\author{E. Ascasíbar}
\affiliation{\ciemat}

\author{TJ-II Team}
\thanks{J.A. Alonso et al 2024 Nucl. Fusion 64 112018}

\date{\today}

\begin{abstract}
	A synthetic Mirnov coils diagnostic for non-axisymmetric magnetic configurations is presented and used to study the capabilities of the poloidal array of single-axis coils and the two helical arrays of tri-axial coils installed in the TJ-II stellarator.
	This tool integrates the plasma currents induced by Alfvén-like perturbations of the electric potential inside the plasma and provides the induced magnetic field oscillations anywhere outside of it.
	The simulated signals can then be analyzed in the same manner as the experimental ones, and a scan on the radial position and width of the potential perturbation is conducted to find the limiting values that produce identifiable signals.
	We find, not surprisingly, that core-localized modes are indistinguishable from one another; and that the identification of low-n, low-m modes is often subject to off-by-one errors.
	We also determine the optimal polarization basis in which to analyze the tri-axial coils signals and address the diagnostic performance when resolving components of gap modes.
	Additionally, selected cases have been analyzed with a simplified plasma response model, showing that plasma shielding of the mode currents may further deteriorate the accuracy of the mode identification method.
	We conclude with the analysis of an experimental case taken from the TJ-II database to illustrate the usefulness of the diagnostic.
\end{abstract}

\maketitle

\section{Introduction} \label{sec:introduction}

The characterization and control of fast-particle driven Alfvén eigenmodes (AEs) is a key issue for burning plasma operation \cite{heidbrinkBasicPhysics2008}.
AEs have been shown to trigger fast-particle transport \cite{duongLossEnergetic1993} and reduce heating efficiency \cite{wongReviewAlfven1999} in fusion devices, therefore deteriorating performance and endangering plasma-facing components.
On the other hand, externally triggered AEs have also been proposed as a plasma exhaust mechanism \cite{kolesnichenkoAlfvenInstabilities2002} which could prove very advantageous in a power plant.
For those reasons, the validation of AE physics models is indispensable for the informed design of future reactors.

The experimental determination of the spatial structure of AEs through the identification of its poloidal (m) and toroidal (n) mode numbers is an integral part of this task, as it allows for theory-experiment comparisons.
Of the set of diagnostics that provide information on mode numbers, Mirnov coil arrays distributed around the plasma column\cite{mirnovProbeMethod1965} are one of the more widely used.
Typically, tokamaks currently in operation like DIII-D \cite{straitMagneticDiagnostic2006}, ASDEX Upgrade \cite{horvathExperimentalInvestigation2016, schittenhelmAnalysisCoupled1997} or TCV \cite{moretMagneticMeasurements1998} have a large number of poloidal Mirnov coils and a much smaller number of toroidal ones, that are usually enough for low-\(n\) modes\cite{minkToroidalMode2016}.
An exception to this trend is MAST-U and its OMAHA poloidal array \cite{holeHighResolution2009} with only 10 coils that are optimized to minimize the spatial aliasing of the measured signal and thus are able to resolve high mode numbers with a comparatively sparse arrangement of sensors.
Compared to stellarators, the toroidal and poloidal mode numbers of the most common modes are more easily identified in tokamaks, with the poloidal number being the most problematic.

The non-axisymmetry of stellarators, however, makes mode identification much more challenging in these devices, since gaps in the shear Alfvén continuum, produced also by helical couplings, lead to modes with different toroidal and poloidal mode numbers that evolve with the same frequency \cite{kolesnichenkoAlfvenInstabilities2002}.
The coil arrangements are also more diverse in stellarators.
LHD has a 16-sensor helical array, along with a 6-sensor toroidal array of tri-axial coils \cite{sakakibaraMagneticMeasurements2010}.
In W7-X there are several poloidal arrays \cite{rahbarniaAlfvenicFluctuations2021, endlerEngineeringDesign2015}, both open and closed, and a smaller number of coils distributed toroidally, for a total of 125 sensors that measure fluctuations in the poloidal direction.
H-1NF had a 16-sensor helical array of tri-axial coils \cite{haskeyMultichannelMagnetic2013} and two poloidal arrays \cite{prettyDataMining2009}.

An in-depth analysis on the limitations of this type of diagnostic for determining both toroidal and poloidal structure of high frequency AEs in non-axisymmetric configurations, in particular modes excited within HAE (Helical Alfvén Eigenmode) gaps, has not been carried out to date.
A recent study by Büschel et al. \cite{buschelSyntheticMirnov2024} presents a synthetic diagnostic to explore the performance of the poloidal arrays of W7-X using different spatial distributions of coils and several magnetic configurations, integrating the perturbed field at the plasma edge using the virtual casing method implemented in the \texttt{EXTENDER\_P} code \cite{drevlakPIESFree2005}.
Some more work has been done in tokamaks: in TCV a study was conducted focusing on the modeling of uncertainties introduced by coil hardware on an ideal mode excited at the coil locations \cite{testaDevelopmentAlgorithms2023} and, in a recent work \cite{bohlsenBayesianFormulation2023}, the response of the Mirnov coils to a distribution of currents was modeled to serve as a forward function for a Bayesian inference-based tomographic reconstruction diagnostic, addressing the possibility of including the plasma response to the current perturbation as well.

In this paper, the arrays of Mirnov coils installed in the TJ-II stellarator have been used for this purpose, as they provide good poloidal and toroidal \cite{ascasibarMeasurementsMagnetic2022} coverage of the device, with a coil arrangement that is similar to the one installed in H-1NF.
NBI (Neutral Beam Injection) driven Alfvén eigenmodes have been extensively studied in TJ-II experiments and previous work focused on the determination of the poloidal mode number was done using SVD (Singular Value Decomposition) \cite{jimenez-gomezAnalysisMagnetohydrodynamic2007, jimenez-gomezAlfvenEigenmodes2011} or spatiotemporal Fourier transforms \cite{vanmilligenMHDMode2012}, without specifically addressing the limitations of the diagnostic.
The work focuses on determining the amplitude of the magnetic field oscillations at the locations of the Mirnov coils produced by an Alfvén wave in the plasma.
Using a new synthetic diagnostic developed for this purpose, an arbitrary potential perturbation, consistent with the structure of the shear Alfvén waves that can be destabilized in the device, is applied to model these oscillations.
We then take the synthetic signal measured by each coil and, using signal analysis techniques such as the 3D Lomb Periodogram, we reconstruct the mode structure and compare the original mode numbers with those obtained from the analysis.
With this diagnostic, we have explored the capabilities of the different sets of Mirnov coils installed in the TJ-II stellarator, performing a scan on the parameters that define the spatial structure of the potential perturbation.
Note also that modelling HAE or TAE (Toroidal Alfvén Eigenmode) coupled modes in the real 3D geometry can provide information on the radial structure of the potential perturbations and their expected low/high field side asymmetries in amplitude, which have been measured in TJ-II using Heavy Ion Beam Probes (HIBP) \cite{melnikovAlfvenEigenmode2012}.
For cases with toroidal mode number \(n=0\), a simplified model of the plasma response has been included in the synthetic diagnostic to investigate the impact of plasma shielding on the mode currents and on the poloidal mode structure as seen by the Mirnov array.

The physics underlying the AEs model used in the synthetic diagnostic and the plasma response calculation is described in appendix \ref{app:model}.
Section \ref{sec:exp_setup} describes the spatial distribution of the different sets of Mirnov coils installed inside the device.
One of them, which measures only one component of the field at each position, is indicated for the measurement of the poloidal mode number while the other two are helical sets of triaxial coils measuring the three components of the magnetic field at each position.
Section \ref{sec:mode_analysis} describes the mode number analysis technique applied to both synthetic and real signals, and section \ref{sec:simul} describes the results of the simulations and addresses the performance of the arrays taking into account aspects such as mode position in the plasma or magnetic field polarization at the measurement locations, which is related to mode polarization in the plasma in a non-trivial way due to the highly three-dimensional structure of the equilibrium field.
Section \ref{sec:experimental_results} showcases the analysis of a discharge of TJ-II, where the lessons learned are applied to experimental data.
Our conclusions and further discussion are held in section \ref{sec:conclusions}.

\section{Experimental set-up} \label{sec:exp_setup}

TJ-II is a medium (\(R_0 =\) 1.5 m, \(a \leq\) 0.22 m, \( V\leq\) 1 m\(^{\text{3}}\)) four-period (\(N_\text{fp}=4\)) heliac stellarator, with magnetic field on axis of \(B_0 = \) 0.95 T.
Low density plasmas heated by a combination of NBI and ECRH (Electron Cyclotron Resonant Heating) are an excellent testing ground for investigating NBI driven Alfvén wave excitation due to their high content of fast ions \cite{cappaStabilityAnalysis2021, melnikovAlfvenEigenmode2012}.
Most of the experiments carried out to date have used the so-called standard magnetic configuration and therefore the performance analysis with the synthetic code will only be done in this configuration.
The magnetic equilibrium was calculated using the \texttt{VMEC} code \cite{hirshmanSteepestdescentMoment1983}, and transformed to Boozer coordinates (\(s, \theta, \varphi\)) using the \texttt{BOOZ\_XFORM} code \cite{sanchezBallooningStability2000}.

\begin{figure}[h] \centering
    \includegraphics[width=\columnwidth]{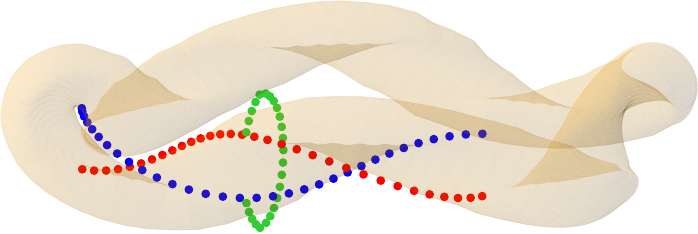}
    \caption{
        Coil locations around the plasma column.
        Poloidal array shown in green, and helical array in blue (upper sub-array) and red (lower sub-array).
    }
    \label{fig:coil_location_plasma}
\end{figure}

In figure \ref{fig:coil_location_plasma}, the coil locations relative to the Last Closed Flux Surface (LCFS) are shown.
The poloidal array (green) consists of 25 coils that measure the changes in magnetic field in the approximate poloidal direction, and covers approximately 270 degrees.
The helical array (blue and red) is made up of two symmetrical sub-arrays that wrap around the central and helical main field conductors (not shown in the figure) for a full period of the device.
Each sub-array consists of 32 tri-axial sensors that allow us to determine the three-dimensional time evolution of the magnetic flux at their center.

\begin{figure}[h] \centering
    \includegraphics[width=\columnwidth]{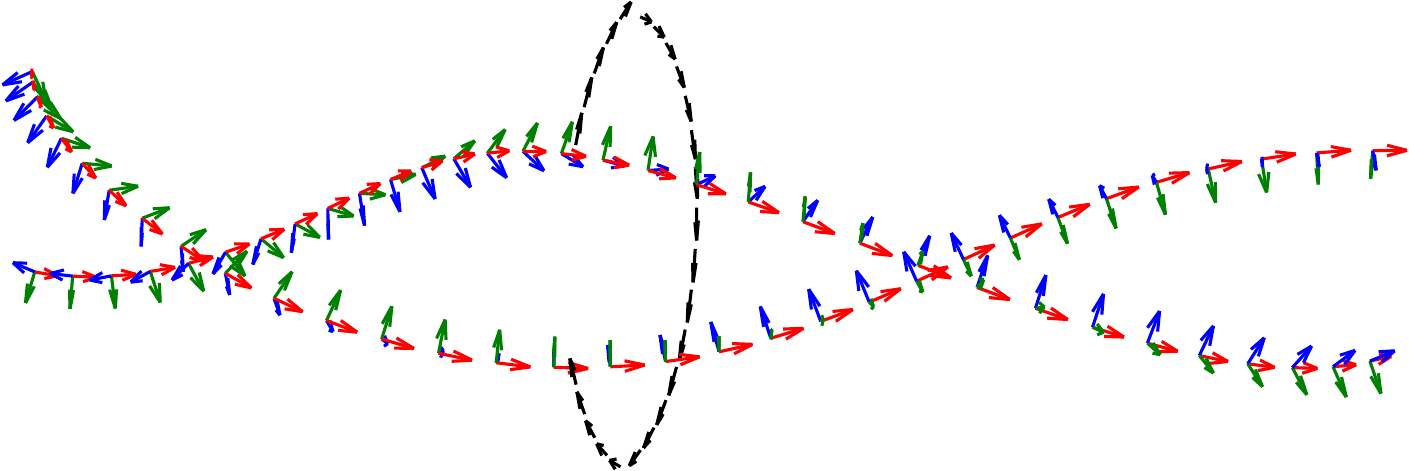}
    \caption{Orientation vectors normal to the detection planes of the helical array tri-axial coils (red, green and deep blue) and the coils of the poloidal array (black).}
    \label{fig:coil_orientation}
\end{figure}

Figure \ref{fig:coil_orientation} depicts the design orientations of the coils.
By construction, the uncertainty in the orientation of the poloidal array coils is low, as they are perfectly aligned inside a rigid metallic tube.
The tri-axial sensors, on the other hand, are located inside a semi-rigid corrugated tube that had to be twisted for its installation inside the vacuum vessel.
The alignment of each tri-axial set of coils inside the tube is more sensitive to positioning errors and a calibration was necessary to determine their true orientations.
Once these orientations are known, the signal can be projected over arbitrary directions in software, as the coils in each sensor are orthogonal.
A comprehensive description of the helical array, its characteristics and the related calibration experiments can be found in ref. \citenum{ascasibarMeasurementsMagnetic2022}.
For the numerical study of the diagnostic performance we have assumed ideal coils distributed according to the position and orientation of the original design, thus avoiding adding an extra layer of complexity.
However, the calibration is essential for the analysis of the experimental data and has been taken into account in the analysis presented in section \ref{sec:experimental_results}.

\section{Mode number analysis} \label{sec:mode_analysis}

\begin{figure}[h] \centering
	\includegraphics[]{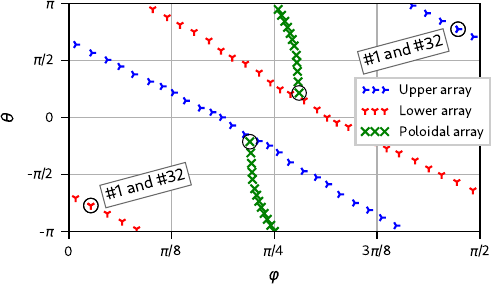}
	\caption{Boozer angles evaluated at the closest point on the LCFS to each coil for the poloidal array (green markers) and the upper (blue) and lower (red) helical sub-arrays.}
	\label{fig:boozer_angles}
\end{figure}

The Lomb-Scargle periodogram \cite{lombLeastsquaresFrequency1976, scargleStudiesAstronomical1982} is widely used in astrophysics to identify periodicities in non-evenly-spaced observation data.
In plasma physics, a 3D generalization was introduced by Zegenhagen\cite{zegenhagenAnalysisAlfven2006} to conduct mode analysis with non-equispaced coils in W7-AS.
The Lomb-Scargle periodogram fits the data \(y_{ij}\) to the sinusoidal model
\begin{equation} \label{eq:lpmodel}
	y_{ij} + \epsilon_{ij} = a \cos(\alpha_{ij}) + b \sin(\alpha_{ij}) \,,
\end{equation}
where \(\epsilon\) is noise, assumed to be white, \(\alpha_{ij} \equiv m\theta_j + n\varphi_j - \omega t_i\) is the phase of the perturbation with mode numbers \(m\) and \(n\) and frequency \(\omega\), with \(i\) being the time label and \(j\) the coil label.
In this model, we assume that both the functional forms of the phase term of the field perturbation outside the plasma and the phase term of the Alfvén mode itself are the same. Since magnetic coordinates are not defined outside the LCFS, we map the outer spatial position of each coil to that of the closest position in the plasma.
Therefore, the coil angles \(\varphi_j\) and \(\theta_j\) are, as in ref. \citenum{zegenhagenAnalysisAlfven2006}, the Boozer angles of the closest point in the LCFS to the coil.
In figure \ref{fig:boozer_angles} they are shown for the standard configuration.

The periodogram is given by
\begin{widetext}
	\begin{equation}
		P_\text{LS}(\omega, n, m) = \dfrac{1}{YY}
		\left(
		\dfrac{\left[\sum_{ij} y_{ij} \cos(\alpha_{ij} - \tau)\right]^2}{\sum_{ij} \cos^2 (\alpha_{ij} - \tau)}
		+
		\dfrac{\left[\sum_{ij} y_{ij} \sin(\alpha_{ij} - \tau)\right]^2}{\sum_{ij} \sin^2 (\alpha_{ij} - \tau)}
		\right) \,,
	\end{equation}
	\label{eq:LombP}
\end{widetext}
where \(\text{YY} = \sum_{ij} y_{ij}^2\) and \(\tau\) is a phase shift term given by:
\begin{equation}\label{eq:tau_definition}
	\tan 2 \tau = \dfrac{\sum_{ij} \sin 2\alpha_{ij}}{\sum_{ij} \cos 2\alpha_{ij}} \,.
\end{equation}

For a more in-depth discussion on this phase shift term, that differs from the one given in \cite{zegenhagenAnalysisAlfven2006}, see appendix \ref{app:tau_lomb}.
With this definition, the periodogram is normalized, giving results in the range \([0,1]\), with \(P_\text{LS} = 0\) corresponding to zero agreement with the measured signals (simulations in this case) and \(P_\text{LS}=1\) corresponding to perfect agreement.
When applied to experimental data, the analysis starts by finding the frequency \(f_0\) of the detected eigenmode from an spectrogram obtained either via the Short-Time Fast Fourier Transform (STFFT) or the DMUSIC (Damped Multiple Signal Classification) method \cite{kleiberModernMethods2021}.
At this time, the frequency is selected manually, but mode following algorithms \cite{vazmendesBroadbandAlfvenic2023, heinrichCharacteristicsAlfvenic2024} can be readily applied for an automated analysis once the method has been shown to be sound.
Then, a scan is performed over the relevant \(n\) and \(m\) mode numbers, that is, the ones that can be resolved given the distance between coils.
The results are plotted as a 2D colormap of \(P_\text{LS}(n, m)\).
For our present purposes, mode frequency and mode numbers are inputs for the synthetic modelling and the potential perturbation associated with the mode is given by equations \ref{eq:pert_potential} and \ref{eq:mode_radial}.

\begin{figure}[h] \centering
	\includegraphics[]{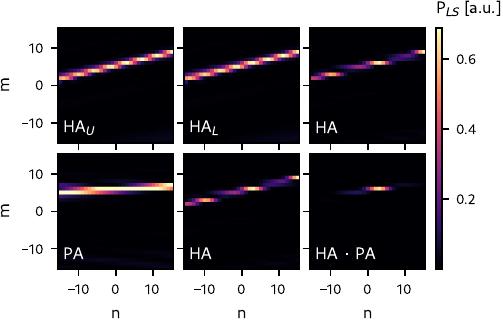}
	\caption{Example Lomb periodograms for the upper (HA\(_\text{U}\)), and lower (HA\(_\text{L}\)) helical sub-arrays, and both of them combined (HA) on the top row; and on the bottom row the poloidal (PA), the combined helical sub-arrays again (HA) for clarity, and the product of the two periodograms (PA\(\cdot\)HA), all for a \(m=6\), \(n=3\) synthetic mode.}
	\label{fig:lomb_example_both}
\end{figure}

The signals measured by the poloidal array have a non-negligible dependency on the toroidal mode number since the array is not located at a constant toroidal magnetic angle, as seen in figure \ref{fig:boozer_angles}.
On the other hand, since the helical sub-arrays follow a straight line in magnetic coordinates, approximately given by
\begin{equation}
	\theta (\varphi) = -N_\text{fp} \varphi + \theta_0 \,,
\end{equation}
the measured signals therefore depend both on the toroidal and the poloidal mode numbers.
This link between poloidal and toroidal angles limits the identifiable mode numbers, introducing false positives in a phenomenon akin to aliasing.
The phase difference between signals measured by two coils separated \(\Delta \varphi\) in the toroidal direction will be:
\begin{equation}
	\delta = (n - N_\text{fp}m) \Delta \varphi = l \Delta \varphi \,,
\end{equation}
where \(l\) is an integer.
For a constant value of \(l\), all the modes that satisfy \(n - N_\text{fp}m = l\) will suffer aliasing and will appear indistinguishable using a single sub-array.
Having two helical sub-arrays that combine measurements at two poloidal locations for each toroidal plane helps mitigate this problem.
This can be seen in the top row of figure \ref{fig:lomb_example_both}, where the Lomb periodogram of each helical sub-array for a simulated mode is depicted.

The periodograms of each sub-array individually are virtually identical, a stripe with slope \(1/N_\text{fp}\), but the extra poloidal information given by the sub-array separation in the poloidal direction is enough to partially remove the uncertainty.
Only after adding the measurements of the poloidal array can we arrive at an unambiguous determination of both toroidal and poloidal mode numbers.
This can be seen best in the bottom row of figure \ref{fig:lomb_example_both}, that shows Lomb periodograms for the poloidal array (PA panel), the two helical arrays, combined (HA panel), and the product of the previous two.
There, the dependency on the toroidal angle \(\varphi\) of the poloidal array shows up as a slight slope in the mode identification of the PA panel, that otherwise (if the array had \(\varphi=\text{constant}\)) would appear as a horizontal stripe.
Similarly, in the HA panel, a diagonal stripe appears, with slope \(1/N_\text{fp}\) as discussed before.
Having both sub-arrays provides enough information to discard some of the modes, and that is the reason why the stripe is not continuous.
The most robust way to perform the mode identification given these limitations in the coil arrangement has proved to be taking the product of the PA and HA periodograms and finding its maximum.
To our knowledge, this is the first time that this method has been used.

As the amplitude of the detected perturbation depends on both the mode numbers and the distance between the coil and the radial position and extension of the perturbation, the normalization of the measured signals has been found to improve the robustness of the mode identification.
For instance, a low intensity signal at some of the coil locations is interpreted by the periodogram as a minimum in the spatial structure of the mode, so false positives can be introduced this way.

\section{Simulations} \label{sec:simul}

We divide this section into two main subsections, one of them dedicated to the simulations with a single synthetic mode and the other to the simulations dealing with coupled modes.
In the first subsection the essential aspects to be taken into account are first illustrated using the poloidal array.
We discuss the treatment of the mode polarization, that is necessary for the mode analysis using the helical arrays.
In the second subsection, where we study coupled mode simulations with both arrays, our goal is to investigate the ability of the diagnostic to resolve different pairs of physically relevant coupled modes.
To account for arbitrary mode coupling, the synthetic diagnostic considers potential perturbations (see appendix \ref{app:model}) of the form
\begin{equation} \label{eq:pert_potential}
	\delta \phi (s, \theta, \varphi, t) = \sum_{mn} \delta \phi_{mn}^\omega (s) e^{i(m\theta + n\varphi - \omega t)} \,,
\end{equation}
where \(\delta \phi_{mn}^\omega \in \mathbb{C}\) is the radial profile of the \(m, n\) mode and \(s=\sqrt{\psi}\) is the radial coordinate used by the VMEC code, being \(\psi\) the normalized toroidal magnetic flux.
As we will only consider linear coupling, the code has been designed so that several modes can be simulated at the same time with arbitrary phase differences and intensities given by the arguments of \(\delta\phi_{mn}^{\omega}\).

The plasma equilibrium must be discretized to carry out the simulations, and a suitably fine grid is mandatory for accurate results.
Convergence studies have shown that, for TJ-II, \(n_s \times n_\theta \times n_\varphi = 100 \times 150 \times 1000\) is enough to get robust results.
The radial derivatives are taken using five-points finite differences, while the {\(\theta\)} and {\(\varphi\)} derivatives are found using Fast Fourier Transforms (FFT), exploiting the flux-surface periodicity of the modes.
The thermal \(\beta\) is rather low in the plasmas that typically exhibit AEs activity in TJ-II, and therefore a vacuum equilibrium is taken for all the studied cases.

The spatial structure of the electrostatic potential perturbation is given by equation \ref{eq:pert_potential}, and the radial profile \mbox{\(\delta \phi_{mn}^\omega\)} is taken to be Gaussian:
\begin{equation} \label{eq:mode_radial}
	\delta \phi_{mn}^\omega (s)
	=
	A_{mn}^\omega \exp \left[ -\dfrac{(s - s_0)^2 }{\sigma} \right] \,,
\end{equation}
with \mbox{\(A_{mn}^\omega \in \mathbb{C}\).}
The current is calculated from the vector potential perturbation using equation \ref{eq:dAmpereA}, which we rewrite here for convenience:

\begin{equation}
	\delta \vb{J}_\text{ext} = - \dfrac{1}{\mu_0} \curl \curl \delta A_\parallel \vb{b}_0 \,.
\end{equation}

\subsection{Single mode simulations} \label{sec:single_mode_simul}

\begin{figure}[h] \centering
	\includegraphics[]{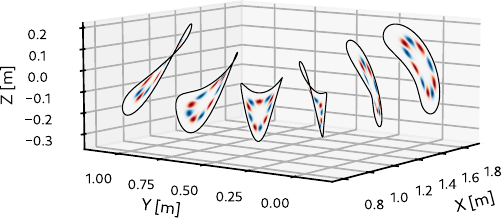}
	\caption{Potential perturbation in six toroidal (\(\varphi = \) constant) cuts along a half-period of the device for a mode with \(m=5\), \(n=7\), \(s_0 = 0.3\) and \(\sigma=0.015\).}
	\label{fig:potentials_3d}
\end{figure}

We start by running the synthetic diagnostic code for single mode perturbations defined by their mode numbers \(n\), \(m\) and their frequency \(\omega\).
In figure \ref{fig:potentials_3d} the potential perturbation for a single \(m=5\), \(n=7\), \(s_0 = 0.3\), \(\sigma=0.015\) mode is shown for several toroidal cuts at constant \(\varphi\) in a half period of the device.

\begin{figure}[h] \centering
	\includegraphics[]{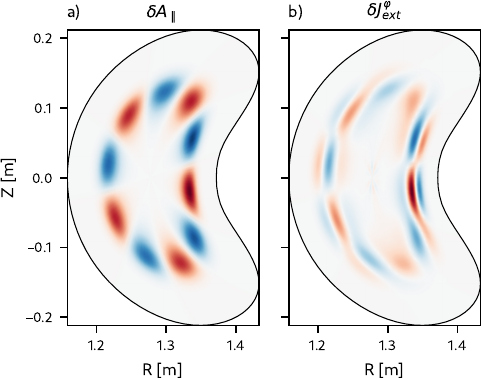}
	\caption{
		2D projection of the plasma cross-section at constant toroidal boozer angle \(\varphi=\pi/4\), showing the parallel component of the perturbed magnetic vector potential (\(\delta A_\parallel\)) in (a) and the contravariant \(\varphi\) component of the current perturbation (\(\delta J_{\text{ext}}^\varphi\)) in (b), for the potential perturbation shown in figure \ref{fig:potentials_3d}.
	}
	\label{fig:jcurr_single}
\end{figure}

Figure \ref{fig:jcurr_single} shows the parallel component of the vector potential and the component along \(\vb{e}_\varphi\) of the induced current associated to the potential pertubation shown in figure \ref{fig:potentials_3d}.
The structure of the current perturbation is notably more complex than that of its potential, which is to be expected since it arises from the double curl of the vector potential.
Also, from this expression we see that narrower modes will have higher current densities, as the slopes will be more steep and thus the derivatives will be larger.

For each  pair of selected mode numbers \((m, n)\), a scan has been conducted over the position of the radial maximum of the perturbed potential, \(s_0\), and its width, parametrized by \(\sigma\).
The resulting simulated perturbations have then been analized with the Lomb periodogram and the dominant mode numbers (\(m_\text{out}, n_\text{out}\)) obtained with the synthetic diagnostic have been compared with the input mode numbers (\(m_\text{in}, n_\text{in}\)).
In the single mode simulations, the mode numbers have been taken arbitrarily and have no direct relation to the actual mode numbers that can be excited in this particular magnetic equilibrium.

\subsubsection{Poloidal array}

We begin by evaluating the ability of the poloidal array to identify poloidal mode numbers.
For that, we must take into account that the signal measured by the probes is \(\delta \dot{B}_\text{p} = \delta \dot{\vb{B}} \cdot \vb{P}\), where \(\vb{P}\) is the vector perpendicular to the coil winding plane (see figure \ref{fig:coil_orientation}), and lies in the RZ plane.

\begin{figure}[h] \centering
	\includegraphics[]{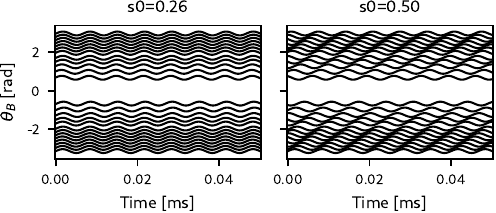}
	\caption{
		Normalized synthetic signals measured by the poloidal array for an \(m=6\), \(n=0\) mode.
		On the left, when the mode is core-localized (\(s_0=0.26\)) and on the right, when a more external location is taken (\(s_0 = 0.5\)).
		The signals have been re-scaled and are spaced in the y-axis proportionally to their associated Boozer angle.
	}
	\label{fig:poloidal_signals}
\end{figure}

In figure \ref{fig:poloidal_signals}, the simulated signals obtained by applying the Biot-Savart integral (equation \ref{eq:biot_savart_deltaphi_appendix}) to the potential of an \(m=6\), \(n=0\) mode are shown, with the maximum of the potential perturbation taken at two different radial positions, a central one (\(s_0 = 0.26\)) and a more external one (\(s_0=0.50\)).
The signals are normalized both for clarity and to avoid distance-dependent effects, and spaced proportionally to the coil separation in poloidal boozer angle.
For this case we have taken modes oscillating at a frequency of 150 kHz.
The differences in the inter-coil phases of the signals are very noticeable, with the core-localized one displaying a very small phase shift between the coils in the array.
The more external mode, however, shows evident phase differences that, when studied with the Lomb periodogram, allow for the correct identification of the mode number.
The phase information of the rotating mode structure appears to be lost when the mode is located deep in the plasma core.

\begin{figure}[h] \centering
	\includegraphics[]{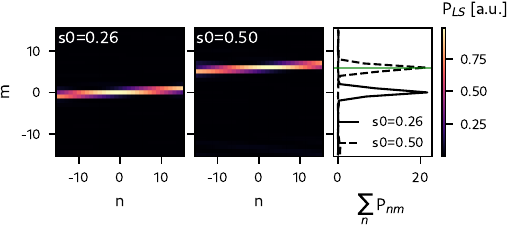}
	\caption{Lomb periodograms applied to the signals shown in figure \ref{fig:poloidal_signals}.
		The outermost mode is correctly identified, while the innermost one is not.
		On the right, the sum over \(n\) of the periodogram values, with the input mode number shown as a green horizontal line
	}
	\label{fig:poloidal_lombs}
\end{figure}

Figure \ref{fig:poloidal_lombs} depicts the Lomb periodograms applied to the signals shown in figure \ref{fig:poloidal_signals}.
On the right of the figure, the sum over \(n\) of the periodogram results is plotted.
The identified poloidal mode number is the one that corresponds with the maximum of this sum.
This method has proven to be the most robust to automatically identify a single \(m\) for the poloidal array, rather than trying to find the maximum over the periodogram results.
The latter approach regularly suffers from off-by-one errors due to the slight slope of the periodogram that was discussed in section \ref{sec:mode_analysis}, as there is a dependence on the toroidal angle \(\varphi\) in the poloidal array.
As expected from the signals, the core localized mode cannot be correctly identified, being instead mistaken for a \(m=0\) mode.
The outermost mode, on the other hand, is correctly identified by the periodogram.

This outcome must be taken into account when analising experimental data since the identification of the core-localized mode number invariably fails.
This is not a limitation of the analysis technique, but rather a well-known limitation of this type of diagnostic that cannot be avoided.
The structure of core-localized modes cannot be experimentally characterized by measurements in the periphery, as is the case for magnetic diagnostics.
For instance, observations in tokamaks \cite{sharapovMonitoringAlfven2004} have shown that magnetic fluctuation diagnostics are indeed less sensitive to core-localized modes and therefore, care must be taken into not trying to analise a mode outside of the diagnostic's operational limits.
For that, diagnostics such as HIBP \cite{melnikov2DDistributions2022}, that can provide potential perturbation profiles, or tomography of soft X-ray (SXR) measurements, proportional to the plasma fluctuations structure \cite{drevalDeterminationPoloidal2021}, are the ideal candidate to complement the analysis.

\begin{figure}[t] \centering
	\includegraphics[]{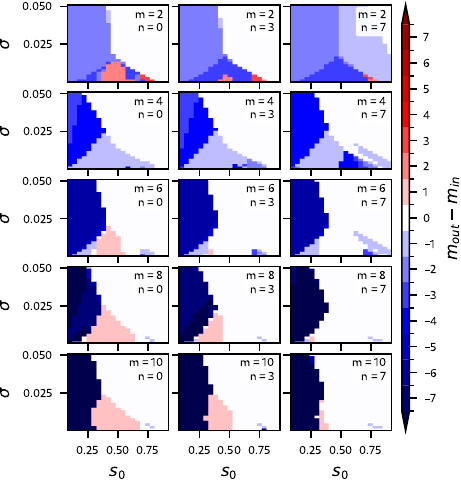}
	\caption{
		Differences between simulated (\(m_\text{in}\)) and measured (\(m_\text{out}\)) poloidal mode numbers for different radial locations and widths of the modes.
		All coils in the poloidal array have been used.
		In white, the points where both mode numbers coincide.
	}
	\label{fig:scan_poloidal_summax}
\end{figure}

\begin{figure}[h] \centering
	\includegraphics[]{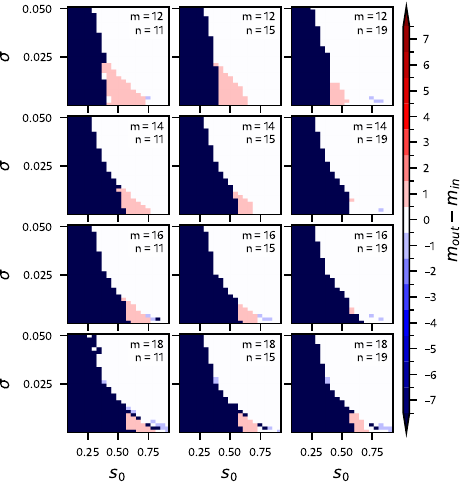}
	\caption{
		Same as figure \ref{fig:scan_poloidal_summax} for high-\(m\) modes.
	}
	\label{fig:scan_highmodes_pol}
\end{figure}

To map these operational limits, a series of scans in both poloidal and toroidal mode number, perturbation width, and radial location have been carried out.
The pattern of core modes losing phase information is common, as we can see in figure \ref{fig:scan_poloidal_summax}, where the difference between the simulated mode number (\(m_\text{in}\)) and the measured mode number (\(m_\text{out}\)) is represented.
In this figure, we can also see that the performance for low-m modes is slightly worse than for higher-m modes, presenting phase information loss for modes further out from the core and more off-by-one errors.
The poloidal array (figure \ref{fig:coil_location_plasma}, green) does not cover the full 360 degrees because the plasma column is very close to the vacuum chamber in the zone of highest indentation of the magnetic surfaces, and it is this lack of coverage that reduces its accuracy for low-m modes.
When studying higher-m modes, as in figure \ref{fig:scan_highmodes_pol}, the performance remains reasonably good for very high mode numbers, even above the nyquist mode number
\begin{equation}
	m_\text{nyq} \simeq \dfrac{1}{2} \dfrac{360 \cdot 25}{270} \simeq 16 \,.
\end{equation}

This is due to the ability of the Lomb periodogram to take advantage of the slightly uneven spacing of the coils in boozer coordinates \cite{vanderplasUnderstandingLombScargle2018}.
For narrower modes, however, the phase information loss appears for more outer modes beginning at \(m\sim 14\).
High $n$ mode numbers have been used in figure \ref{fig:scan_highmodes_pol} because the simultaneous detection of high mode numbers in both the poloidal and toroidal direction should be the most challenging situation for the diagnostic, and so that a meaningful comparison of the performance after the addition of the helical array in section \ref{ssec:helical_array} can be carried out.
We will see later that this is not the case, and high mode numbers can be resolved remarkably well.

\subsubsection{Wave polarization}
\label{sec:wave_polarization}

Now we turn to study the polarization of the detected perturbation.
Wave polarization is a property of the wave that specifies the spatial orientation of the electric and magnetic field as the wave evolves in time. It carries information on the amplitude of the different components of the wave and the phase shifts between them in a given basis of polarization vectors \cite{jacksonClassicalElectrodynamics1999}.
The locus of the \mbox{\(\delta \vb{B}\)} vector during one wave period is known as the polarization ellipse and lies in a plane whose orientation in space depends precisely on the polarization properties of the wave.
Inside the plasma, the transverse polarization of shear Alfvén waves makes the perturbed magnetic field perpendicular to the equilibrium field; but this does not hold outside, where the field perturbation detected at a given spatial position is given by an integral over the whole plasma volume.
Even though the poloidal array only measures \(\delta \dot{\vb{B}}\) along a predetermined direction, the synthetic diagnostic provides the full 3D signal, opening up the possibility of an optimization study on the best possible orientation of a set of single axis coils for mode identification.
Finding the right polarization base to guarantee an optimal result is an essential part of the analysis.

\begin{figure}[h] \centering
	\includegraphics[]{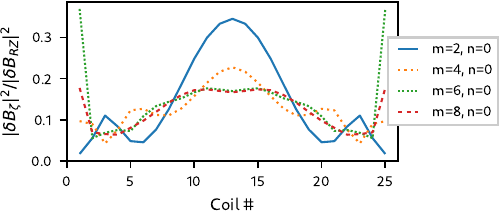}
	\caption{
		Ratio of squared azimuthal \mbox{\(|\delta  B_\zeta|\)} to radial-axial \mbox{\(|\delta B_\text{RZ}|\)} amplitudes detected by the poloidal coils for \(n=0\) modes with different poloidal mode numbers.
	}
	\label{fig:poloidal_ratio_parperp}
\end{figure}

\begin{figure}[t] \centering
	\includegraphics[]{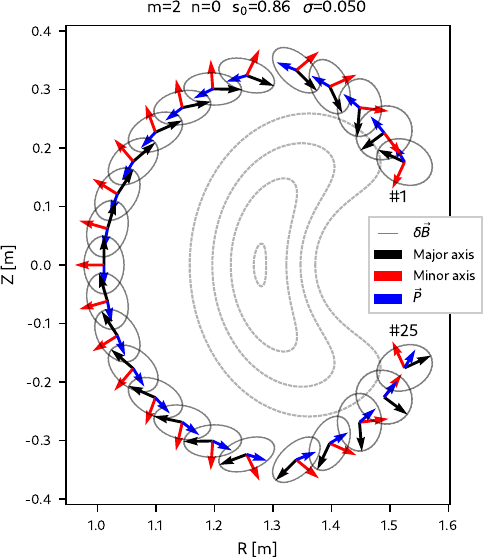}
	\caption{2D projection \mbox{\(\delta B_\text{RZ} (t)\)} of the general \(\delta{\vb{B}}(t)\) polarization ellipses for an \(m=2\), \(n=0\) mode.
		The perpendicular vectors to each coil winding plane (\(\vb{P}\), blue arrows) and the vectors indicating the major (black arrows) and minor (red arrows) axis of the polarization ellipses at each location are shown.}
	\label{fig:polarization_ellipse_poloidal}
\end{figure}

As a first approach, and considering the geometry of the poloidal array, we restrict ourselves to the projection of the general 3D polarization ellipse on the RZ plane, trying to find out if there is an optimal orientation for the coils, always keeping \mbox{\(\vb{P}\)} in the RZ plane.
In figure \ref{fig:poloidal_ratio_parperp} we show the ratio between the amplitude of the oscillating field in the azimuthal direction (\mbox{\(|\delta B_\zeta|\)}) to that in the RZ plane (\mbox{\(|\delta  B_\text{RZ}|\)}).
A very low value of this ratio indicates that the oscillating field lies mostly in the RZ plane, so we can expect a better performance of the diagnostic in such cases.
This ratio depends on both the mode numbers and the radial position of the modes, and there are instances where it is not small.
Figure \ref{fig:polarization_ellipse_poloidal} shows the (normalized) projection of the \mbox{\(\delta  B_\text{RZ} (t)\)} polarization ellipse over the RZ plane with their major and minor axes and the vectors \(\vb{P}\), that determine the coil orientations, for an \(m=2\), \(n=0\) mode.
Note the changes in angle between \(\vb{P}\) and the major axis of the ellipse, that become more pronounced on the rightmost coils.

\begin{figure}[h] \centering
	\includegraphics[]{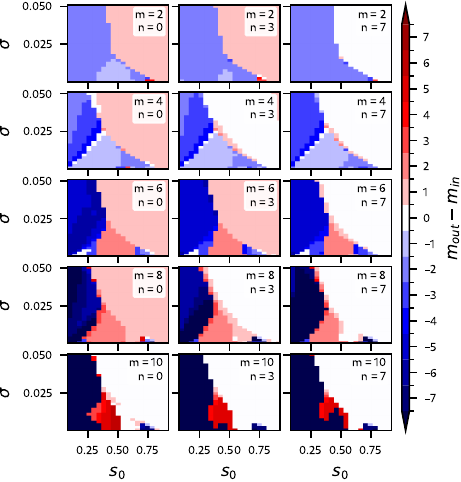}
	\caption{Same as figure \ref{fig:scan_poloidal_summax} but using only the innermost coils (\(R<1.3\) m) in the poloidal array.
	}
	\label{fig:scan_poloidal_middlecoils}
\end{figure}

It might seem, after inspection of the result shown in figure \ref{fig:polarization_ellipse_poloidal}, that mode identification accuracy could be improved by performing the periodogram analysis on the components of the field perturbation in the local polarization basis defined by the major and minor axes of the projected ellipses.
That is, rotating \mbox{\(\vb{P}\)} so that it aligns with the major/minor axis of the ellipse at each coil.
However, this method has been tested and it has proved to not be effective, and projecting over the design orientation given by \(\vb{P}\) (blue arrows in figure \ref{fig:polarization_ellipse_poloidal}) yields much better results.
Furthermore, as the polarization of the measured signal is mode-dependent, one cannot design an uni-axial coil array that performs well for any mode, and the experimental use of the aforementioned procedure is not feasible without having tri-axial sensors.

Going back to figure \ref{fig:polarization_ellipse_poloidal}, we see that the coils where the angle between \(\vb{P}\) and the major axis of the ellipse changes more rapidly, when moving from coil to coil, are the ones that are closest to the plasma.
Figure \ref{fig:scan_poloidal_middlecoils} shows the results of the mode analysis with the rightmost five coils removed on each side, hoping for a decrease in the distortion of the mode structure and an improvement of the detection accuracy.
We observe that the detection accuracy does not improve, and off-by-one errors appear more frequently.
This could be caused by a lack of angular coverage of the plasma column, but that would mainly affect the detection of low-\(m\) modes (\(m\leq3\)), which does not seem to be the case.
Alternatively, the issue may lie in the coils furthest from the plasma, as the part of the plasma that contributes to the Biot-Savart integral with significant intensity increases with coil distance (as the contribution of a volume element is proportional to \(1/r^2\)).
On the other hand, the detection limits (the radial position and mode width where the mode identification is not possible) do not change significantly.

Finding a suitable polarization basis is
essential for the helical array, as having tri-axial coils forces us to choose an orientation along which to project the temporal derivative of the magnetic field vector in order to conduct the analysis.
The design orientations of the coils shown in \ref{fig:coil_orientation} are a good starting point, as they project the signal into, roughly, its radial, toroidal, and poloidal components; yielding acceptable results.

However, we can do better \cite{haskeyExperimenttheoryComparison2015} by taking into account the magnetic configuration.
We can take the direction of the magnetic field at the closest point on the LCFS for each coil, that for typical low-pressure, low-current TJ-II plasmas is very close to the (normalized) dual \(\varphi\) basis vector\cite{dhaeseleerFluxCoordinates1991}:
\begin{equation}
	\hat{\vb{b}} \simeq \dfrac{\grad \varphi}{\norm{\grad \varphi}} = \hat{\vb{e}}^\varphi \,;
\end{equation}
the normal to the LCFS, that corresponds with the normalized dual \(s\) basis vector:
\begin{equation}
	\hat{\vb{n}} \equiv \dfrac{\grad s}{\norm{\grad s}} = \hat{\vb{e}}^s \,;
\end{equation}
and their binormal, that turns out to be the normalized tangent \(\theta\) basis vector\cite{dhaeseleerFluxCoordinates1991}:
\begin{equation}
	\hat{\vb{b}} \times \hat{\vb{n}} \simeq \hat{\vb{e}}_\theta \,.
\end{equation}

\begin{figure}[h] \centering
	\includegraphics[]{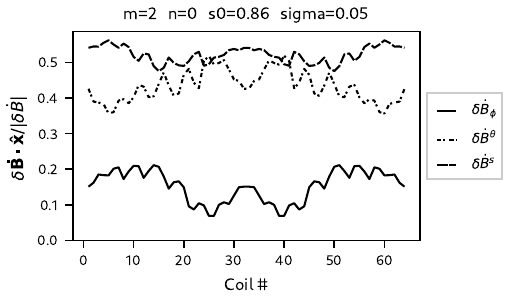}
	\caption{Relative magnitude of \(\delta\dot{\vb{B}}\) projected on the normalized basis \(\left\{ \hat{\vb{e}}^\varphi, \hat{\vb{e}}_\theta, \hat{\vb{e}}_s \right\}\) calculated at the closest point on the LCFS for the helical array.}
	\label{fig:relative_amplitudes_sub_base}
\end{figure}
As shown in figure \ref{fig:relative_amplitudes_sub_base}, and as expected due to the nature of the simulated perturbation, the magnitude of the signal along \(\hat{\vb{e}}^\varphi\) is consistently much lower, making \( \{ \hat{\vb{e}}_\theta, \hat{\vb{e}}^s \}\) a good polarization base for mode number analysis.
This is also consistent with the conclusions of the previous section for which \(\vb{P}\), which is mainly directed along \(\hat{\vb{e}}_\theta\), proved to be the best choice.

\begin{figure}[h] \centering
	\includegraphics[]{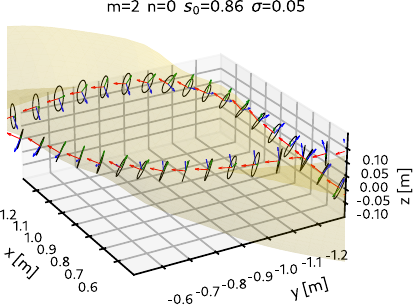}
	\caption{
		3D polarization ellipses of the field perturbation at the coil positions (black) and normalized basis vectors \(\hat{\vb{e}}^s\) (blue), \(\hat{\vb{e}}_\theta\) (green), \(\hat{\vb{e}}^\varphi\) (red) for a \(m=2\), \(n=0\) mode.
	}
	\label{fig:helical_pol_3d}
\end{figure}

Figure \ref{fig:helical_pol_3d} shows the (normalized) three-dimensional perturbation of a \(m=2\), \(n=0\) mode along with the normalized basis vectors for a section of the helical array.
The polarization ellipses are represented in the figure.
As seen most clearly in the lower coils, these are very well aligned with the \(\left\{ \hat{\vb{e}}_\theta, \hat{\vb{e}}^s \right\}\) basis vectors, although it is also clear that the projection over \( \hat{\vb{e}}^\varphi\) is not zero.

\subsubsection{Helical array}
\label{ssec:helical_array}

\begin{figure}[b] \centering
	\includegraphics[]{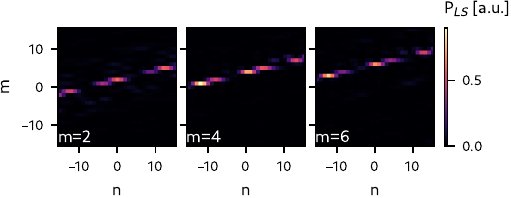}
	\caption{Lomb periodograms for three different modes (\(m=2\), \(m=4\), and \mbox{\(m=6\)}) calculated only with the \(\hat{\vb{e}}_\theta\) projection of the signal, using only the coils in the helical array.}
	\label{fig:helical_lombs}
\end{figure}

\begin{figure}[b] \centering
	\includegraphics[]{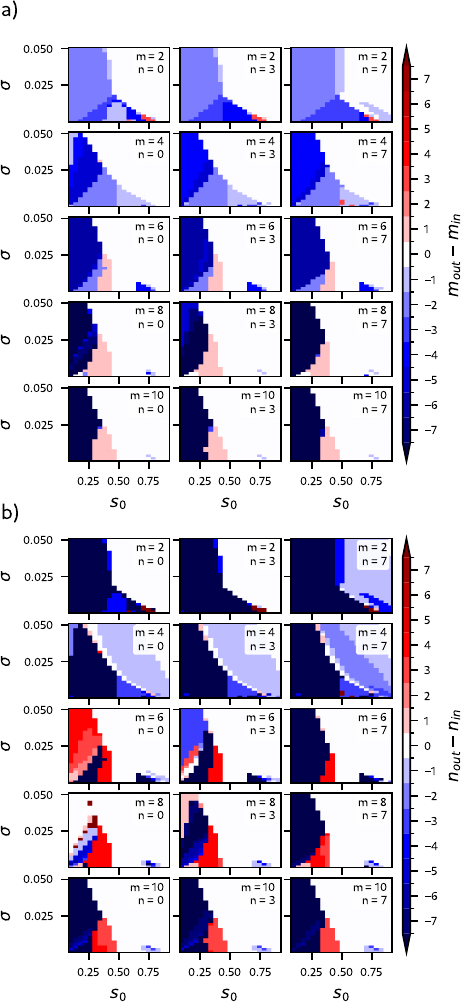}
	\caption{
		Differences between input (\(m_\text{in}, n_\text{in}\)) and output (\(m_\text{out}, n_\text{out}\)) poloidal (a) and toroidal (b) mode numbers for the scan on radial maximum and width of a single mode.
		For the mode analysis, both the poloidal and helical (projecting over \(\hat{\vb{e}}_\theta\)) arrays have been used.
		In white, the points where both mode numbers coincide.}
	\label{fig:ms_ns_tor_pol_merged_lowmn}
\end{figure}

In figure \ref{fig:helical_lombs}, the Lomb periodograms of the \(\hat{\vb{e}}_\theta\) projection of the signal for three modes with different poloidal mode number \(m\) are shown, only for the coils in the helical array.
Following the discussion in section \ref{sec:mode_analysis}, we may find the value of \(l\) as a function of the input mode numbers \(m_\text{in}, n_\text{in}\) and we can get the equation for the stripe of aliased modes:
\begin{equation}
	m_\text{al} = \dfrac{n_\text{al}}{N_\text{fp}} + l= \dfrac{n_\text{al}}{N_\text{fp}}+\dfrac{m_\text{in} N_\text{fp} - n_\text{in} }{N_\text{fp}} \,.
\end{equation}

\begin{figure}[t] \centering
	\includegraphics[]{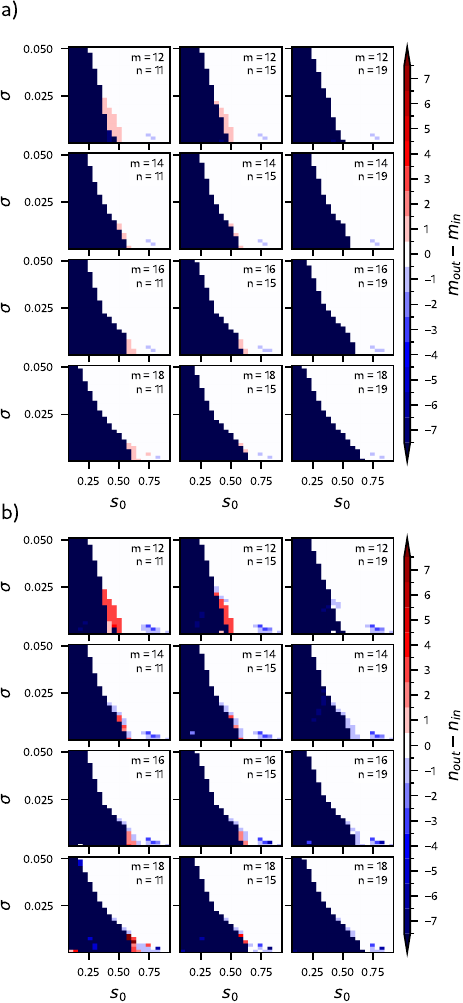}
	\caption{
		Same as figure \ref{fig:ms_ns_tor_pol_merged_lowmn} for high mode numbers
		Differences between input (\(m_\text{in}, n_\text{in}\)) and output (\(m_\text{out}, n_\text{out}\)) poloidal (a) and toroidal (b) mode numbers.
	}
	\label{fig:ms_ns_tor_pol_merged_highmn}
\end{figure}

Turning back to figure \ref{fig:helical_lombs}, we see that as the poloidal mode number increases the stripe of possible modes moves up as expected.
From this figure, it is noteworthy that the \(n, m\) pair with the highest intensity returned by the periodogram does not in general coincide with the input mode numbers \(n_\text{in}, m_\text{in}\).
This shows that, for such an arrangement, the poloidal array is essential for the identification of both the poloidal and the toroidal mode numbers.
The employed polarization basis has demonstrated superior performance compared to all other studied configurations.
As explained in section \ref{sec:mode_analysis}, the detected mode number is found multiplying the poloidal and helical periodograms and finding the maximum.
This allows us to identify both poloidal (\(m\)) and toroidal (\(n\)) mode numbers.

Figure {\ref{fig:ms_ns_tor_pol_merged_lowmn}} shows the difference between the simulated and identified mode numbers using both the helical and poloidal arrays for the low-\(m\), low-\(n\) case.
Qualitatively, the behaviour remains similar to the case with only the poloidal array, which suggests that the latter is limiting the performance of the former.
For low-\(m\), low-\(n\) modes the mode identification performance deteriorates because the arrays do not close in on themselves and only cover part of the device.
Due to the link between toroidal and poloidal angles of the helical array, discussed above, the identification errors in toroidal mode number usually occur in steps of four, meaning that for a mode \(n_\text{r}\) the most frequently identified modes will satisfy \(n_\text{id} = n_\text{r} + 4k\), where \(k \in \mathbb{Z}\).
Figure {\ref{fig:ms_ns_tor_pol_merged_highmn}} shows the same result for the high-\(m\), high-\(n\) cases.

\FloatBarrier

\subsection{Coupled modes simulations} \label{sec:mmode}

Generally, Alfvén eigenmodes are destabilized in gaps in the Alfvén continuum that emerge from periodicities in the Alfvén velocity produced in turn by periodicities of the magnetic field structure.
While, generally, perturbations in the plasma equilibrium are strongly damped if their frequencies lie on the continuum, such damping does not happen (or is much weaker) if their frequency falls in a gap.
Such gaps can originate either from zeros in the radial variation of the continuum frequency (\(\partial \omega / \partial r = 0\)) or from frequency crossings of counter-propagating waves \cite{heidbrinkBasicPhysics2008}.
Similarly to what happens in tokamaks, where modes of the same \(n\) and different \(m\) can evolve jointly at the same frequency within the TAE gap, the existence of helicity gaps (HAE) in stellarators allows for complex radially extended and weakly damped structures characterized by modes with different values of \(n\) and \(m\).
For instance, in the case of an HAE\(_{\mu\nu}\) gap created by the appropriate components in the Fourier expansions of the magnetic field and the metric coefficients, modes with \(n_1\) and \(m_1\) can couple with modes with \(n_2=n_1+\nu N_\text{fp}\) and \(m_2=m_1+\mu\).

\begin{figure}[h] \centering
	\includegraphics[]{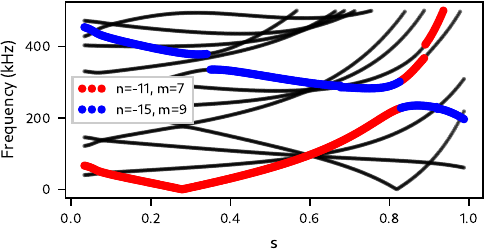}
	\caption{Shear alfvén continua calculated with \texttt{STELLGAP} (\(n=1\) family only), with relevant gap modes highlighted.}
	\label{fig:stellgap}
\end{figure}

We have used the code \texttt{STELLGAP}\cite{spongShearAlfven2003} to find these gaps in the continuum for a experimentally relevant equilibrium (calculated with \texttt{VMEC}) and then simulated the synthetic signal that a combination of both modes would produce.
In figure \ref{fig:stellgap}, the Alfvén continuum for the \(n=1\) mode family is represented.
The relevant gap occurs at \(s \sim 0.8\), and is caused by the coupling of a \(m_1=7, n_1=-11\) and a \(m_2=9, n_2=-15\) mode.
The modes differ in both \(m\) and \(n\), and the difference in values (\(m_2 - m_1 = 2\)) and (\(n_2 - n_1 = 4 = 1N_\text{fp}\)), allow us to classify it as a \(\text{HAE}_{21}\).
Once the relevant modes have been identified, our interest lies in the performance of the diagnostic and in its ability to properly resolve the two modes.
Although linear stability simulations\cite{cappaStabilityAnalysis2021} show that generally one of the modes is dominant, we have taken here the same amplitude for each of the coupled modes.

\begin{figure}[t] \centering
	\includegraphics[]{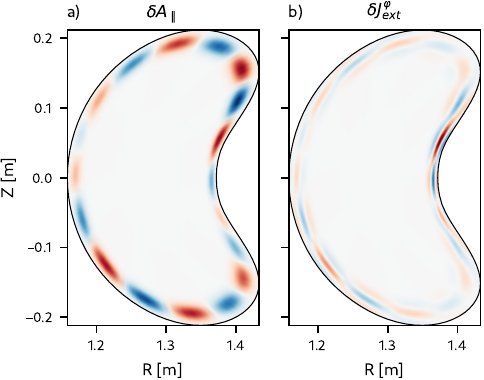}
	\caption{Same as figure \ref{fig:jcurr_single} but now calculated for a superposition of modes consistent with figure \ref{fig:stellgap}, with mode numbers \( m_1=7,\ n_1=-11 \) and \( m_2=9,\ n_2=-15 \).
	}
	\label{fig:jcurr_double}
\end{figure}
To investigate the mode resolving power of the diagnostic, we have performed a scan in radial separation of the modes, keeping the gaussian profiles used so far, and also introducing an arbitrary phase difference between both modes.
The potential and current perturbations created by such combination of modes are shown in figure \ref{fig:jcurr_double}.
There, the potential shows nine maxima and minima, and also a \(m=2\) modulation in the amplitude of the perturbations, in accordance with the difference between the poloidal mode numbers of the modes.

\begin{figure*}[h] \centering
	\includegraphics[]{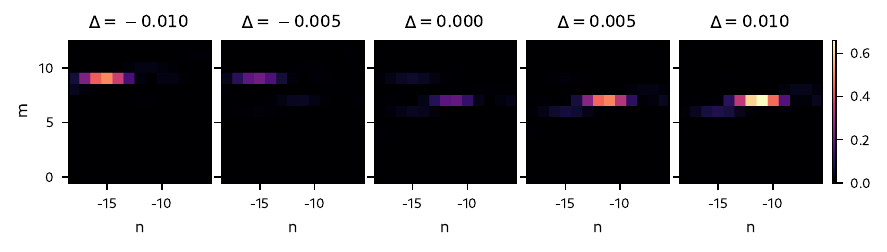}
	\caption{Lomb periodograms calculated for the HAE$_{21}$ gap mode.
		Several cases with different radial separation between maxima of the two coupled modes are analyzed.}
	\label{fig:mapas_multi_modes_scan_delta}
\end{figure*}

For this scan, we introduce a quantity \(\Delta\) that measures the separation between the radial maximum of the modes, so that, if \(s_0 = 0.8\), one mode will be centered around \(s_1 = s_0 - \Delta\) while the other will be centered around \(s_2 = s_0 + \Delta\).
Performing a scan over this quantity, and analising the results with the Lomb periodogram, we get figure \ref{fig:mapas_multi_modes_scan_delta}.
There, we see that the outermost mode is the only one that is identified, and that the periodogram intensity, that is an indication of the confidence in the identification, decreases when the modes overlap the most.
Still, for \(\Delta = 0\), the identified mode is the \(m=7,\ n=-11\) (although a considerably fainter maximum is also present for the \(m=9,\ n=-15\) mode as well).
The analysis shows that the diagnostic cannot resolve an HAE$_{21}$ gap mode into its coupled modes, and that for modes with very close maxima, the component with lowest m (therefore higher signal as was shown in fig. \ref{fig:dists}) will appear with higher intensity in the analysis.
Whether or not this conclusion applies for TAEs (same $n$) or for modes with lower mode numbers is left for future analysis that can also benefit from experimental input.

\onecolumngrid

\phantom{.}
% \FloatBarrier
\twocolumngrid

\section{Experimental results}
\label{sec:experimental_results}

To illustrate the usefulness of the synthetic diagnostic when analyzing real experimental data, we take an example from the TJ-II database.
The chosen shot belongs to a series of experiments aiming at characterizing the influence of Electron Cyclotron Current Drive (ECCD) on the Alfvén spectrum of NBI-heated plasmas \cite{cappaStabilityAnalysis2021, PonsEPS2023}.
The plasma was started with second harmonic ECRH, using two 53.2 GHz gyrotrons, with one of the ECRH beams configured to provide co-ECCD at lower power, helping also with density control.
At \(t=1125\) ms, co-NBI heating is introduced, and one of the gyrotrons is switched off shortly after.

\begin{figure}[h] \centering
	\includegraphics[]{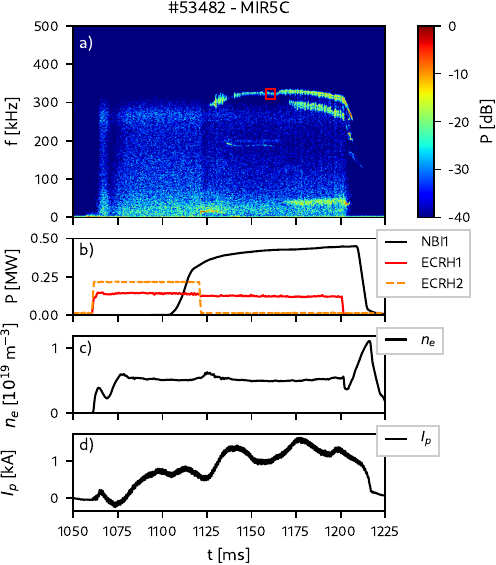}
	\caption{
		Spectrogram of magnetic fluctuations (a), heating scheme (b), line integrated density (c), and plasma current (d)
		A red rectangle (top) illustrates the time-frequency range used to make figure \ref{fig:dmusic}.
	}
	\label{fig:exp_spgram}
\end{figure}

The spectrogram of the magnetic fluctuations detected with one of the Mirnov coils is shown in figure \ref{fig:exp_spgram}-a).
Below, the heating scheme (fig. \ref{fig:exp_spgram}-b), the line-integrated density (fig. \ref{fig:exp_spgram}-c) and the plasma current (fig. \ref{fig:exp_spgram}-d) of the discharge are pictured.
The spectrogram shows that, a few milliseconds after turning on the neutral beam, a mode with \(f_1\sim 320\) kHz, probably TAE or HAE appears.
This mode presents a slight chirping character, mainly during the final phase of the discharge, where another chirping mode with frequency \(f_2 \sim 290\) kHz appears, at the same time than a low-frequency non-Alfvénic mode.

The changes in the MHD spectrum can be explained by changes in the rotational transform due to the stabilization of the NBI-driven part of the plasma current.
This has been discussed elsewhere \cite{CappaIAEA2023} and the physics interpretation as well as a precise mode identification analysis lies outside of the scope of this paper.

\begin{figure}[h] \centering
	\includegraphics[]{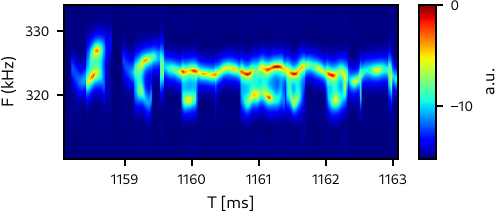}
	\caption{DMUSIC spectrogram of selected range.}
	\label{fig:dmusic}
\end{figure}

This shot has been selected because the mode frequency changes minimally during its lifetime, thanks to a very good density control.
This is helpful when conducting the analysis, as the Lomb periodogram is only valid for a constant mode frequency.
In modes with relatively slow frequency changes, however, this limitation can be overcome by a piecewise analysis.
In any case, a constant frequency is clearer for the illustrative purposes of this section.
Figure \ref{fig:dmusic} shows a spectrogram of a suitable time interval of the discharge, calculated with the DMUSIC algorithm \cite{kleiberModernMethods2021}, where the narrow, constant-frequency nature of the mode can be appreciated.

\begin{figure}[h] \centering
	\includegraphics[]{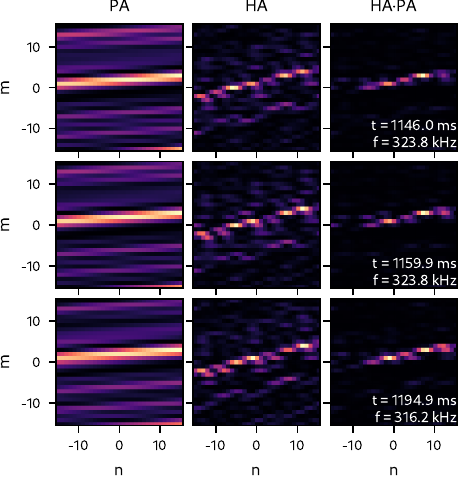}
	\caption{Lomb periodograms of the \(f_1 \sim 320\) kHz mode, taken at three distinct times (rows). The left column shows the periodogram calculated using only the poloidal array signals. The middle column presents the same calculation but with the helical array. The rightmost column shows the product of the previous two.}
	\label{fig:lomb_experimental}
\end{figure}

The results of the analysis in three distinct intervals of the discharge of the \(f_1 \sim 320\) kHz mode are shown in figure \ref{fig:lomb_experimental}.
The results are very reproducible, although they lack the clarity of the results obtained using synthetic signals and an unambiguous identification of the dominant mode number is difficult, as the maximum in the product of periodograms (shown in the rightmost column) changes between \(m=3,\ n=7\) in the first two rows and \(m=1,\ n=-1\) in the last.
This uncertainty can be attributed to the conditions of a real experiment where some of the sensors of the poloidal array, damaged by ECRH stray radiation, were not in operation.

\begin{figure}[b]
	\includegraphics{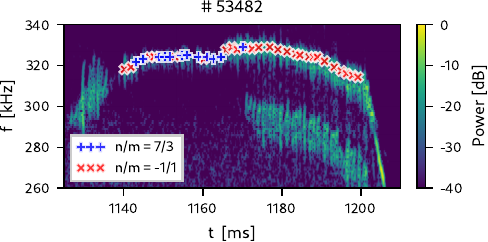}
	\caption{
		Zoomed in spectrogram and identified mode numbers for the \mbox{\(f_1 \sim 315\)} kHz mode.
	}
	\label{fig:lomb_freq_spgram}
\end{figure}

\begin{figure}[h]
	\includegraphics{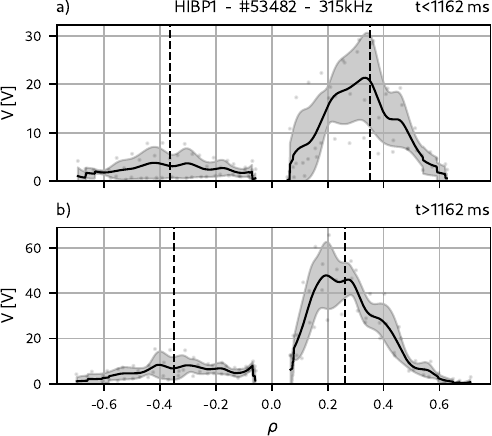}
	\caption{
		Electrostatic perturbed potential profiles measured with HIBP, for \mbox{\(t<1162\)} ms (a) and \mbox{\(t>1162\)} ms (b).
		Vertical dashed lines mark the mean position of the mode during each time interval.
		\mbox{\(\rho<0\)} corresponds to the high field side of the plasma, and \mbox{\(\rho>0\)} to the low field side.
	}
	\label{fig:hibp}
\end{figure}

A time-resolved mode number measurement can be carried out by repeating this process for a series of suitable time intervals.
In figure \ref{fig:lomb_freq_spgram} the measured mode numbers obtained in this way are plotted over the zoomed-in spectrogram. We observe that for \mbox{\(t<1165\)} the mode numbers are most commonly \mbox{\(n/m = 7/3\)}, while for \mbox{\(t>1165\)} the pair of mode numbers \mbox{\(n/m = -1/1\)} are measured.
To understand this result, or at least, to know to what extent we can take it as good, we used a combination of HIBP measurements and results provided by the synthetic diagnostic.
Note that while the density remains constant along the entire length of the shot, the current does not.
The combined effect of the Neutral Beam Current Drive (NBCD) and ECCD, that both induce positive current, results in an increase of the rotational transform which is here the main driver of changes in the spectrum.
The radial profiles of the perturbed electrostatic potential, measured with one of the heavy ion beam probes are shown in figure \ref{fig:hibp} for \mbox{\(t<1162\)} (a, top) and \mbox{\(t>1162\)} (b, bottom).
They exhibit a strong ballooning character typical of gap mode and, most importantly, the maximum of the profile appears to move inward for the second part of the shot.
This is seen more clearly on the low field side (\mbox{\(\rho > 0\)}) since the mode amplitude is larger there.
Taking into account the synthetic diagnostic results regarding mode number errors when mode radial location shifts inward (\ref{fig:ms_ns_tor_pol_merged_lowmn}), the \mbox{\(-1/1\)} mode number measurement is very likely not accurate, corresponding with the unidentifiable core-localised modes that were discussed in the previous section.
This underscores the synthetic diagnostic's ability to effectively discern trustworthy mode number measurements from those that are not.
However, this is not the only possible explanation.
Alternatetively, changes in rotational transform could also explain why the destabilized modes are different, in addition to having different radial positions.
To rule out this possibility, a more comprehensive set of simulations, out of the scope of the paper, is needed to precisely explain the behavior of this eigenmode.

The above example illustrates how the synthetic diagnostic can help in the interpretation of the results but not simulate the measurement itself since it shows that, in the case of gap modes, the various coupled modes cannot be resolved to begin with.
Therefore, in the above case, there is insufficient experimental information available to model the measurement of the coils.

\begin{figure}[h] \centering
	\includegraphics[]{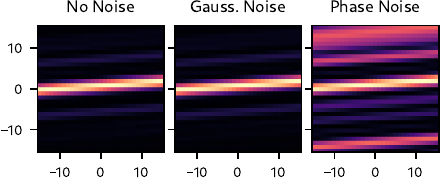}
	\caption{Periodograms of a synthetic \(m=1,\ n=1\) mode, using the poloidal array only. Left, with the noiseless signal. Center, with added gaussian noise. Right, with phase noise.}
	\label{fig:lomb_noise}
\end{figure}

The simulations and analysis conducted up until now have always been performed considering the best-case scenario: an exact positioning of the coils, no instrumental effects on the signal, no electromagnetic noise from the bulk plasma, no eddy currents in the nearby vessel nor in the metallic tubes housing the coils, etc.
When considering experimental data, however, these sources of error must be taken into account.
As a first approach to understand the impact that all these sources of error have on the experimental results, we can explore the effect that numerical noise added to the synthetic signals has on the mode analysis results.
We take, for this excercise, a pair of mode numbers compatible with the ones detected experimentally.
In figure \ref{fig:lomb_noise}, three periodograms made with synthetic data over the experimentally available coils of the poloidal array are shown.
The first one, calculated with no noise, serves as a benchmark.
Comparing with i.e. figure \ref{fig:poloidal_lombs}, we can already see a slight broadening of the band, that is caused by the decrease in the number of available signals.

For the second one, gaussian noise has been added to the signals, according to
\begin{equation}
	y_{ij}^\text{n} = y_{ij} + \epsilon_{ij} \,
\end{equation}
where \(\epsilon_{ij}\) is a random sample from a normal distribution with standard deviation scaled to, in this case, a 15\% of the signal amplitude.
The effects of this random noise are barely noticeable.
We can think of the periodogram as a least-squares fit, and so intuitively this kind of noise should "cancel out" for a long enough time interval.

The third periodogram is the most interesting.
There, a small random phase offset \(\delta_j\) is added to the signal of each coil, via:
\begin{equation}
	y_{ij} = \Re\left\{ y_{ij}^\text{c} e^{i\delta_j} \right\} \,.
\end{equation}
Here, \(y_{ij}^\text{c}\) is the complex output of the synthetic diagnostic.
This random phase offset models both any uncertainties in the spatial position of the coils, that would change their associated Boozer angles; the effects of eddy coils in the vaccuum vessel and nearby plasma-facing components, and the data acquisition system could have on the digitized signal.
These have been observed and corrected in other devices \cite{horvathReducingSystematic2015} but a similar work has not been done yet in TJ-II.
This phase noise significantly degrades the results of the periodogram, but if we assume it is either small (not necessarily the case, see ref. \citenum{horvathReducingSystematic2015} again) or similar for all coils (more likely, as the surrounding environment and data acquisition pipeline are comparable), the mode number can still be identified.

\section{Conclusions and future work} \label{sec:conclusions}

We have outlined the development of a synthetic diagnostic to model the effect of Alfvén-like perturbations on the measurement of ideal Mirnov coil arrays.
The fundamental objective was to study the capability of both the diagnostics and the analysis techniques used for mode number determination.
In this way, we have not only narrowed down the optimal operating ranges of the poloidal and helical arrays of coils installed in the device, but also addressed different ways of analyzing the signals, in particular in terms of the optimal polarization basis.

Besides the expected bad performance for core lozalized modes, synthetic analysis shows a recurring appearance of off-by-one errors in the mode number determination.
This needs to be taken into account when comparing mode number measurements to theoretical predictions.
Most of the modelling has been carried out for modes with single \(m\) and \(n\), as we can expect from GAEs for instance or for gap modes (TAEs or HAEs) with some of its components clearly dominating over the others.
The case in which we have considered an HAE\(_{21}\) gap mode with similar coupled modes amplitudes shows that the diagnostic cannot resolve between both coupled modes and that only one of its components is detected depending, on this case, on the separation between maxima.

Although we have carried out extensive simulations, in terms of mode numbers and mode location mostly, many other parameters have remained unexplored and we can use this tool to complement the analysis of experimental data.
For instance, including more complicated profiles, as the ones measured with HIBP or SXR, or the ones calculated by simulations with MHD codes, is straightforward.
Moreover, we have only used in this work the standard magnetic configuration with no toroidal current.
It is easy to switch to other configurations, taking into account also possible changes in iota that may affect the shear Alfvén spectrum, by recalculating the equilibrium and remapping the coils in Boozer coordinates.

One of the advantages of the tri-axial sensors of the helical arrays is that they allow us to study the polarization of the mode.
The relationship between mode polarization and polarization of the measured radiation at the sensors position is not straightforward and the reconstruction of the mode polarization from the measurements needs prior knowledge of the measured mode numbers and a good measurement of the radial profile of the mode potential\cite{PonsEPS2023} or, failing that, a synthetic database including several combinations of the former.
This will be addressed in the near future using the experimental data presented in\cite{PonsEPS2023}.

Finally, solving the wave equation by Fourier methods and including the response of an homogeneous plasma through its cold dielectric tensor, we have analyzed the plasma response for modes with \mbox{\(n=0\)} and arbitrary mode number \mbox{\(m\)}.
The model is limited in several important aspects, all of them discussed in appendix \ref{app:plasma_response_model}, but it shows that measured mode numbers may differ from the real ones.
Combining this result with the one obtained with the vacuum case, which is basically a Biot-Savart integral, shows that deviations of the order of \mbox{\(\Delta m=\pm 2\)} or larger, even for modes not located in the plasma core, can be expected between measurements and simulations.
Furthermore, this needs to be considered when determining the toroidal mode number, which depends in turn on a good charaterization of the poloidal one, as an error of \mbox{\(\delta m \pm 1\)} leads to errors of \mbox{\(\delta n\pm 4\)}.

Beyond all possible hardware-related considerations or impact of surrounding structures on the measurements (non-ideal coils), it is essential to take all the previous issues into account whenever model validation against experimental results is attempted.
We have taken an experimental case, in which part of the coils of the poloidal array were not working, to demonstrate the use of synthetic diagnostic both for mode number determination including noise modelling and polarization measurements.

The code for this synthetic diagnostic, that can be easily used to conduct similar analysis for other devices, is publicly available in the following GitHub repository: \url{https://github.com/pponsv/synth_mirnov}.

\section*{Acknowledgements} \label{sec:acknowledgements}

The authors are grateful to K. Rahbarnia, R. Kleiber, S. Vaz Mendes, and C. Büschel for the helpful discussions and advice during the initial stages of this work.
This work was partially funded by the Spanish Ministry of Science and Innovation under several contracts, Grant Nos. ENE2013-48109-P, FIS2017-88892-P, FIS2017-85252-R, PID2021-125607NB-I00, and by the European Regional Development Fund (ERDF) “A way of making Europe”.
This work has been carried out within the framework of the EUROfusion Consortium, funded by the European Union via the Euratom Research and Training Programme (Grant Agreement No 101052200 — EUROfusion).
Views and opinions expressed are however those of the author(s) only and do not necessarily reflect those of the European Union or the European Commission.
Neither the European Union nor the European Commission can be held responsible for them.

\appendix
\section{Physical model}
\label{app:model}

The starting equations are Maxwell's equations for a general strongly dispersive non-isotropic medium:
\begin{align}
    \curl\vb{E}+\frac{\partial \vb{B}}{\partial t}
     &
    =0 \label{eq:Faraday}
    \\
    \curl\vb{B}
     &
    =\mu_0\vb{J}_\text{ext}+\frac{1}{c^2}\frac{\partial \vb{E}}{\partial t}+\mu_0\vb{J}(\vb{E}) \,, \label{eq:AmpereG}
\end{align}
where we have separated the external current imposed on the system, i.e. equilibrium current and a perturbation to this equilibrium current (\(\vb{J}_\text{ext}=\vb{J}_{ext,0}+\delta\vb{J}_\text{ext}\)), from the current \(\vb{J}(\vb{E})\) flowing in the plasma in response to the perturbation.
The last two terms on the right of \ref{eq:AmpereG} correspond to the displacement current in ordinary dielectrics.
In a plasma, all charge carriers are free and the distinction between polarization and conduction currents disappears \cite{brambillaKineticTheory1998}, which allows us to write Maxwell equation \ref{eq:AmpereG} as shown.
In the present context, the equilibrium current (\(\vb{J}_{ext,0}\)) stands for all static currents flowing in the plasma and in the device coils and will not be included in the analysis.

\subsection{Vaccum electromagnetics}
\label{app:emvac_model}

In the low frequency quasi-static limit, and disregarding for the moment the plasma response, \ref{eq:AmpereG} becomes
\begin{equation}
    \curl\vb{B}=\mu_0\vb{J}_\text{ext}
    \label{eq:AmpereT}
\end{equation}
This equation, combined with \ref{eq:Faraday}, still allow us to study low frequency standing waves as the ones described by the MHD approximation, in which displacement current plays a negligible role.
Shear Alfvén waves fall into this category and together with momentum balance equation and Ohm's law for the MHD fluid, equation \ref{eq:AmpereT} is taken as the starting point to address the MHD stability problem.
There is a large body of work devoted to MHD stability analysis in tokamak and stellarator configurations (for ideal MHD see \cite{freidbergIdealMHD2014} and references therein).
For our present purposes we will be content to note that the typical structure of the current perturbations, which is formally given by \ref{eq:AmpereT},
\begin{align}
    \delta\vb{J}_\text{ext} & =\frac{1}{\mu_0}\curl\delta\vb{B}      \\
                            & =\frac{1}{\mu_0}\curl\curl\delta\vb{A}
    \label{eq:dAmpereT}
\end{align}
can be easily cast into a useful expression using prior knowledge of the spatial structure of the perturbed electric potential.
For shear Alfvén waves in the ideal MHD limit (\(\delta{E_{\parallel}}=0\)), it can be shown that \cite{vladDynamicsAlfven1999, kolesnichenkoAlfvenEigenmodes2002}
\begin{equation}
    \dfrac{\partial}{\partial t}\delta A_\parallel=-\grad_\parallel\delta\phi.
    \label{eq:apar}
\end{equation}
where \(\delta A_{\parallel}\) is the parallel component of the perturbed vector potential \(\delta\vb{A}\) and \(\delta\phi\) is the perturbation to the electric potential.
Moreover, since shear Alfvén waves can be reasonably described in the compression-less limit (\(\delta B_{\parallel}=0\)) we can take \(\delta\vb{A}_\perp=0\) and write
\begin{equation}
    \delta\vb{J}_\text{ext}=\frac{1}{\mu_0}\curl\curl\delta A_{\parallel}\vb{b}_0.
    \label{eq:dAmpereA}
\end{equation}
where \(\vb{b}_0\equiv\vb{B_0}/B_0\) and \(B_0\) is the equilibrium magnetic field.
It can be shown that equation \ref{eq:AmpereT} is the differential form of the Biot-Savart Law \cite{jacksonClassicalElectrodynamics1999} and thus, in the low frequency limit, the magnetic field created by the perturbation at any point in space, can be calculated as
\begin{equation}\label{eq:BiotS}
    \delta \vb{B}(\vb{r},t)
    =
    \dfrac{\mu_0}{4\pi} \int_\text{V} \text{d}^3 r' \ \dfrac{\delta \vb{J}_\text{ext}(\vb{r}', t) \times \vb{R}}{R^3}
\end{equation}
where \(\vb{R} \equiv \vb{r} - \vb{r}'\).
Note that, in expression \ref{eq:BiotS}, \(\delta \vb{B}(\vb{r},t)\) response to \(\delta \vb{J}_\text{ext}(\vb{r}', t)\) is instantaneous since the quasi-static limit is assumed.
The solution of \ref{eq:AmpereG} in vacuum (i.e. dropping the plasma response term \(\mu_0\vb{J}(\vb{E}))\) is given by the general expression for time-varying current densities in a volume introduced by Jefimenko \cite{jefimenkoSolutionsMaxwell1992}:
\begin{equation}\label{eq:jefimenko}
    \delta \vb{B} (\vb{r}, t)
    =
    \dfrac{\mu_0}{4\pi} \int_V  \text{d}^3 r' \
    \left( \dfrac{\left[\vb{\delta J}\right] \times \vb{R}}{R^3}
    +
    \dfrac{\tfrac{\partial \left[\vb{\delta J}\right] }{\partial t} \times \vb{R} }{c R^2} \right)
\end{equation}
where the square brackets denote evaluation at \(r'\) and retarded times \(t' = t - R/c\).
Griffiths and Heald \cite{griffithsTimedependentGeneralizations1991} showed that if we expand the retarded current around \(t\) in both the first and second terms of equation \ref{eq:jefimenko}, the terms on the first derivative cancel out, and we are left with:
\begin{equation}
    \delta \vb{B} (\vb{r}, t)
    \simeq
    \dfrac{\mu_0}{4\pi} \int_V \text{d}^3 r' \
    \left(
    \vb{\delta J} - \dfrac{1}{2} \dfrac{R^2}{c^2} \dfrac{\partial^2 \vb{\delta J}}{\partial t^2}
    \right)
    \times \dfrac{\vb{R}}{R^3}
\end{equation}
If the currents associated to the perturbation oscillate as \(\vb{J} = \vb{J}_0 e^{i\omega t}\), the Biot-Savart law will be a good approximation as long as:
\begin{equation}
    \left( \dfrac{\omega R}{c} \right)^2 \ll 1
\end{equation}
which is the case for alfvénic instabilities in current experimental devices.

Our interest lies in the time derivative of the magnetic field, which is the physical quantity measured by the sensors.
Thus, taking the time derivative on both sides of \ref{eq:BiotS} and using \ref{eq:dAmpereA} and \ref{eq:apar}, equation \ref{eq:BiotS} becomes
\begin{equation}
    \dfrac{\partial\ }{\partial t} \ \delta \vb{B}
    =
    -\dfrac{1}{4\pi} \int_\text{V} \text{d}^3 r' \ \dfrac{\left(\curl \curl \grad_\parallel\delta\phi\right) \times \vb{R}}{R^3}.
    \label{eq:biot_savart_deltaphi_appendix}
\end{equation}
This integration can be performed numerically and can be quite time-consuming when a very fine mesh is taken to limit numerical errors.
For a given set of modes with spatial periodicity defined by its mode numbers \(m\) and \(n\) and a given constant frequency \(\omega\), we can express the perturbed potential as a Fourier series on the magnetic angles \(\vartheta\) and \(\varphi\) (here taken to be Boozer angles):
\begin{equation}  \label{eq:mode_theta_phi}
    \delta\phi^{\omega}= \sum_{m,n} \delta\phi_{mn}^\omega (s) e^{i\chi} e^{ -i \omega t}
\end{equation}
where \(\chi=(m\vartheta + n\varphi)\) is the spatial phase.
The time dependence can be extracted from the integral in \ref{eq:biot_savart_deltaphi_appendix}, so we end up with a sum of complex integrals of the spatial part of the Fourier series:
\begin{equation}
    \dfrac{\partial\ }{\partial t} \ \delta \vb{B} (\vb{r},t)=
    \sum_{m,n} \vb{I}_{mn}^{\omega}(\vb{r}) e^{-i\omega t}
\end{equation}
where \(\vb{I}_{mn}^{\omega}(\vb{r})\) is the Biot-Savart integral of each mode:
\begin{widetext}
    \begin{equation}
        \vb{I}_{mn}^{\omega}(\vb{r})
        \equiv-\dfrac{1}{4\pi} \int_\text{V} \text{d}^3 r'\dfrac{
            \left\{ \curl \curl \grad_\parallel \left(\delta\phi_{mn}^\omega(s) e^{i\chi}\right)     \right\} \times\vb{R}}{R^3}
    \end{equation}
\end{widetext}
and the integral has to be taken over the plasma volume.
For convenience (no magnetic coordinates are defined outside the plasma volume) the components of the current vector inside the braces, which are first calculated in magnetic coordinates, are transformed to Cartesian coordinates and \(\text{d}^3 r'=ds d\vartheta d\varphi\sqrt{g}\), being \(\sqrt{g}\) the Jacobian of the coordinate transformation.
The advantage of directly taking the electrical potential of a given mode with frequency \(\omega\) is that it allows the data measured by heavy ion beam probes (HIBP) to be used directly in the simulation.
On the other hand, MHD stability codes such as FAR3d or CKA provide directly either \(\delta\phi\) or \(\delta{A}_{\parallel}\).

\subsection{Homogeneous plasma response for n=0 modes}
\label{app:plasma_response_model}

All the results presented in the main body of the paper have been obtained using a Biot-Savart integral in vacuum, without taking into account the plasma response, which is not included in the model described above since we have neglected the displacement currents.
To evaluate the effect of the plasma on the radiated fields we start from the wave equation for \(\vb{E}\), obtained by taking the curl of \ref{eq:Faraday} and using \ref{eq:AmpereG},
\begin{equation}
    \curl\curl\vb{E}+\frac{1}{c^2}\frac{\partial ^2\vb{E}}{\partial t^2}+\mu_0\frac{\partial\vb{J}(\vb{E})}{\partial t}=-\mu_0\frac{\partial\vb{J}_\text{ext}}{\partial t}
    \label{eq:waveeq}
\end{equation}
The constitutive relation \(\vb{J}(\vb{E})\) in a plasma is much more complicated that in ordinary dielectrics since the current induced by \(\vb{E}\) at any point in space and time depends on the previous history of the electric field in the surrounding space, that is
\begin{equation}
    \vb{J}(\vb{r},t)=\int_{-\infty}^t dt'\int \text{d}^3 r'\sigma(\vb{r} - \vb{r}',t-t')\cdot\vb{E}(\vb{r}',t')
\end{equation}

This expression, which is valid only for a homogeneous plasma since the conductivity kernel \(\sigma\) is only a function of the space time distances and not of the specific locations, leads to a tractable model developed in many textbooks on plasma waves (see for instance \cite{brambillaKineticTheory1998, dumontWavesPlasmas2017, stixWavesPlasmas1992}).
By Fourier-Laplace transforming \ref{eq:waveeq} in space and time respectively, it can be shown that the perturbed amplitude \(\delta\vb{E}_{\vb{k},\omega}\) obeys the following equation
\begin{equation}
    \vb{k}\times\vb{k}\times\delta\vb{E}_{\vb{k},\omega}+\frac{\omega^2}{c^2}\epsilon_{\vb{k},\omega}\cdot\delta\vb{E}_{\vb{k},\omega}=-i\omega\mu_0\delta\vb{J}_{\vb{k},\omega}
    \label{eq:waveeq_fourier}
\end{equation}
where
\begin{equation}
    \epsilon_{\vb{k},\omega}=\mathbb{I}+\frac{i\sigma_{\vb{k},\omega}}{\epsilon_0\omega}
    \label{eq:dtensor}
\end{equation}
is the plasma dielectric tensor and \(\sigma_{\vb{k},\omega}\) the conductivity tensor.
The amplitudes \(\delta\vb{J}_{\vb{k},\omega}\) are the Fourier-Laplace transform of the perturbed current source given by \ref{eq:dAmpereA}.
For low frequency waves in the range of hundreds of kilohertz, the cold plasma dielectric tensor is diagonal \cite{brambillaKineticTheory1998} and its components can be written as
\begin{align}
    \epsilon_{xx} & \simeq\frac{\omega^2}{c^2}-\frac{\omega_{pe}^2}{c^2} \\
    \epsilon_{yy} & \simeq\frac{\omega^2}{c^2}+\frac{\omega^2}{v_A^2}    \\
    \epsilon_{zz} & =\epsilon_{yy}
\end{align}
where \(v_A=B_0/\sqrt{\mu_0 m_in_i}\) is the Alfven velocity (disregarding electron mass density) and
\(\omega_{pe}^2=e^2n_e/\epsilon_0m_e\) is the squared electron plasma frequency.
To facilitate the integration with the overall TJ-II reference system, we haven taken \(\mathbf{B}_0=B_0\hat{\vb{x}}\) instead of the usual Stix choice \cite{stixWavesPlasmas1992} that takes the static field \(\vb{B}_0\) along \(z\).
Therefore, components \(\epsilon_{yy}\) and \(\epsilon_{zz}\) account for the behaviour perpendicular to the field while \(\epsilon_{xx}\) does it for the parallel part.
For waves in the \(100-400\) kHz range excited in low density strongly magnetized plasmas with \(B_0\sim 1\) T and \(n_e\sim 10^{19}\), we have \(\omega^2/c^2\sim10^{-6}\), \(\omega^2/v_A^2\sim10^{-2}\) and \(\omega_{pe}^2/c^2\sim10^{5}\).
Using the Einstein summation convention, equation \ref{eq:waveeq_fourier} takes the form
\begin{equation}
    \Lambda^{ij}\delta E_{\vb{k},\omega}^{j}=-i\omega\mu_0\delta J_{\vb{k},\omega}^{i}
    \label{eq:waveeq_indexsum}
\end{equation}
where we have defined the dispersion tensor \(\mathbf{\Lambda}(\mathbf{k,\omega})\) as
\begin{equation}
    \Lambda^{ij}=k^{i}k^{j}-k^2\delta^{ij}+\frac{\omega^2}{c^2}\epsilon^{ij}.
    \label{eq:dis_tensor}
\end{equation}
Provided \(\mathbf{\Lambda}(\mathbf{k,\omega})\) is an invertible matrix, the solution of \ref{eq:waveeq_indexsum} can be obtained as
\begin{equation}
    \delta E_{\vb{k},\omega}^{i}=-i\omega\mu_0 G^{ij}\delta J_{\vb{k},\omega}^{j}
    \label{eq:waveeq_sol}
\end{equation}
being \(\mathbf{G}=\mathbf{\Lambda}^{-1}\) the inverse of \(\mathbf{\Lambda}(\mathbf{k,\omega})\). Taking the Fourier-Laplace transform of \ref{eq:Faraday} and using \ref{eq:waveeq_sol} we obtain the perturbed magnetic field in Fourier space,
\begin{equation}
    \delta\vb{B}_{\vb{k},w}=\frac{\vb{k}\times\delta\vb{E}_{\vb{k},w}}{\omega}.
    \label{eq:waveeq_solB}
\end{equation}
Finally, taking the inverse Fourier transform of \ref{eq:waveeq_solB}, the time evolution of the field in real space is recovered
\begin{equation}
    \delta \vb{B} (\vb{r},t)=\left[\int\delta\vb{B}_{\vb{k},w}\, e^{i\mathbf{k}\cdot\mathbf{r}} d^3\mathbf{k}\right] e^{-i\omega t}
    \label{eq:field_fourier}
\end{equation}

\begin{figure}[b]
    \includegraphics[width=1.0\columnwidth]{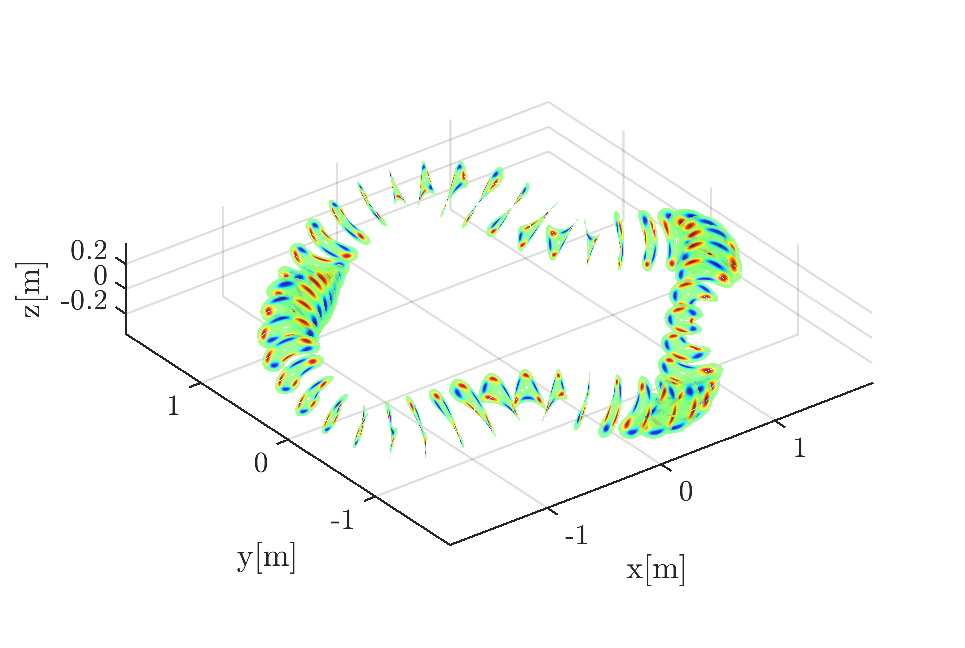}\\%
    \includegraphics[width=0.5\columnwidth]{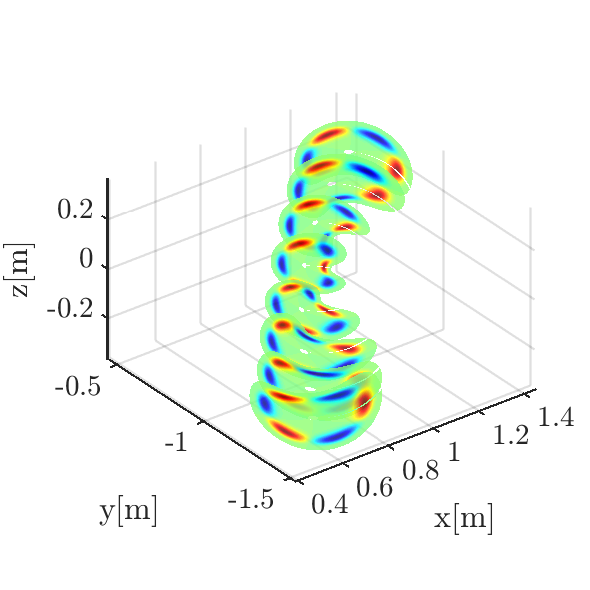}%
    \includegraphics[width=0.5\columnwidth]{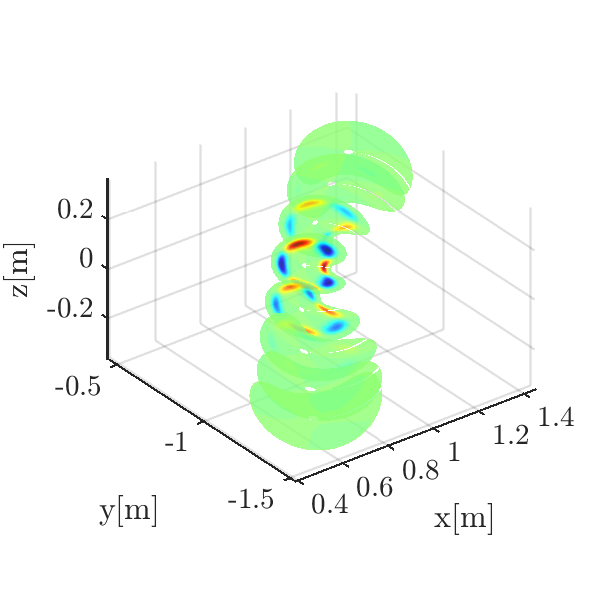}%
    \caption{
        On top, the full \mbox{\(n=0, m=3\)} toroidally extended perturbation, with \mbox{\(s=0.5\)} and \mbox{\(\sigma=0.03\)}, is shown. Bottom left shows the "straight" plasma column section.
        The same perturbation, radially localized around the detection plane of the poloidal array of Mirnov coils is shown on the bottom right.
    }
    \label{fig:mode_local}%
\end{figure}

By evaluating the field at the positions of the coils we can calculate the synthetic signal and account for the effect that the response of a homogeneous plasma of a given density immersed in a constant magnetic field would have on an arbitrary distribution of currents.
Since the typical current distributions we are dealing with in this paper are far from being the ones that could be created in a homogeneous plasma, we will restrict ourselves to an ideal case that nevertheless maintains the essential characteristics of the problem. In order to obtain a meaningful estimate of the plasma response while minimizing the impact of approximating the response of the real non-homogeneous plasma to that of a constant density one with magnetic field directed along the main direction of the real 3D field, we assume a highly localized mode current distribution centered toroidally at the toroidal plane where the poloidal array of Mirnov coils is located.

\begin{figure}[h]
    \includegraphics[width=\columnwidth]{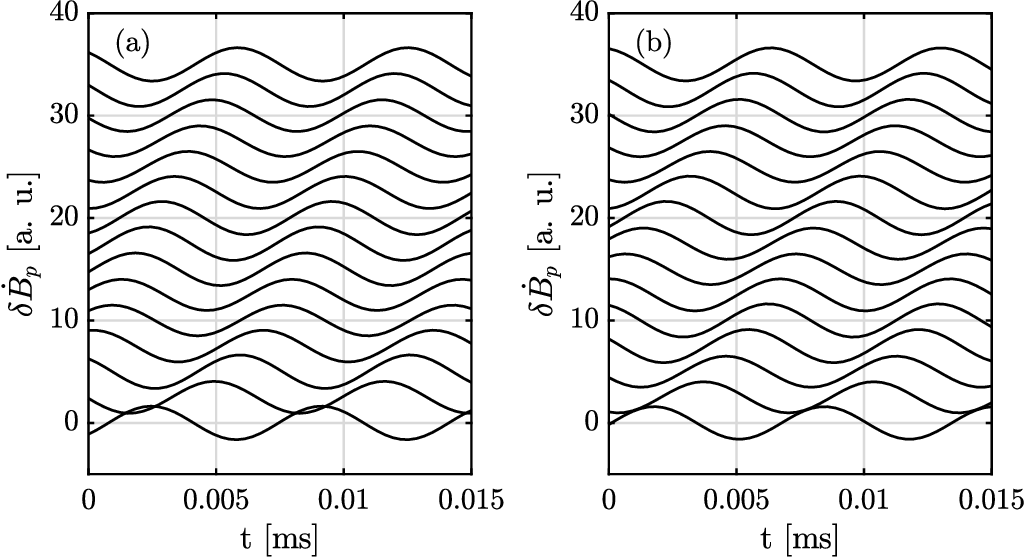}%
    \caption{Synthetic signals generated by the toroidally extended (a) and localized (b) \(n=0, m=3\) perturbation that is shown in figure \ref{fig:mode_local}.
    }
    \label{fig:comp_loc_ext}%
\end{figure}

Both the extended and toroidally localized potential distributions are shown in figure \ref{fig:mode_local}.
This approximation is valid as long as we restrict ourselves to $n=0$ modes. Only in this case the synthetic signals generated by the localized current distribution are a good approximation to the ones generated by a toroidally extended realistic distribution of currents. This is illustrated in figure \ref{fig:comp_loc_ext}, where the corresponding synthetic signals are compared.
Only slight differences appear and the \(m=3\) structure is well preserved in the case of the localized perturbation.

\begin{figure}[h]
    \includegraphics[width=\columnwidth]{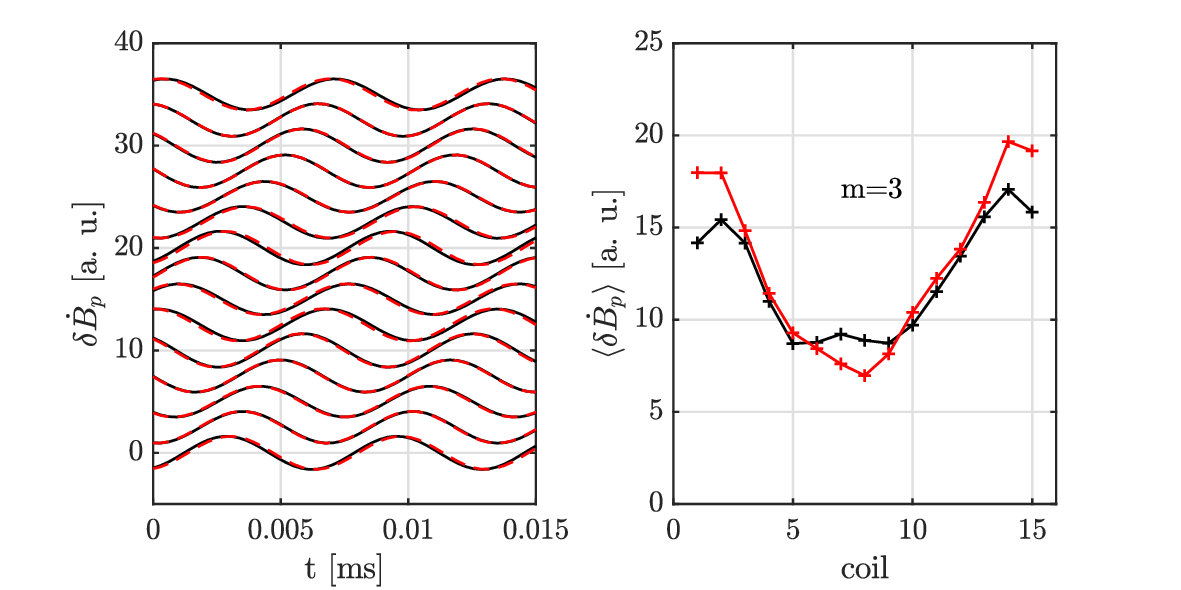}\\
    \includegraphics[width=\columnwidth]{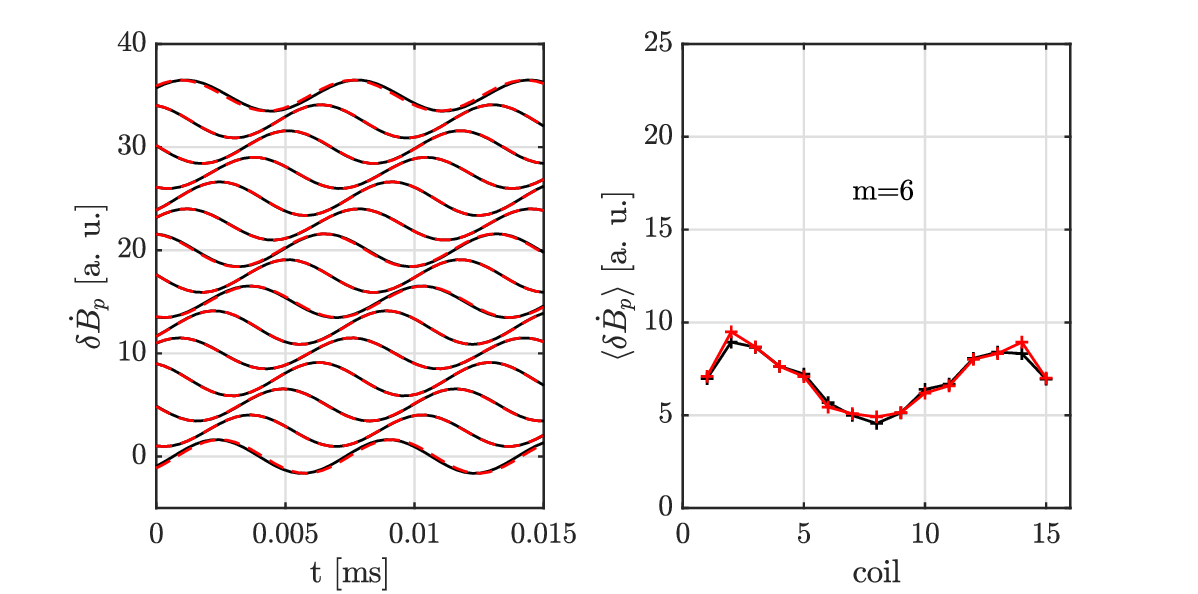}\\
    \includegraphics[width=\columnwidth]{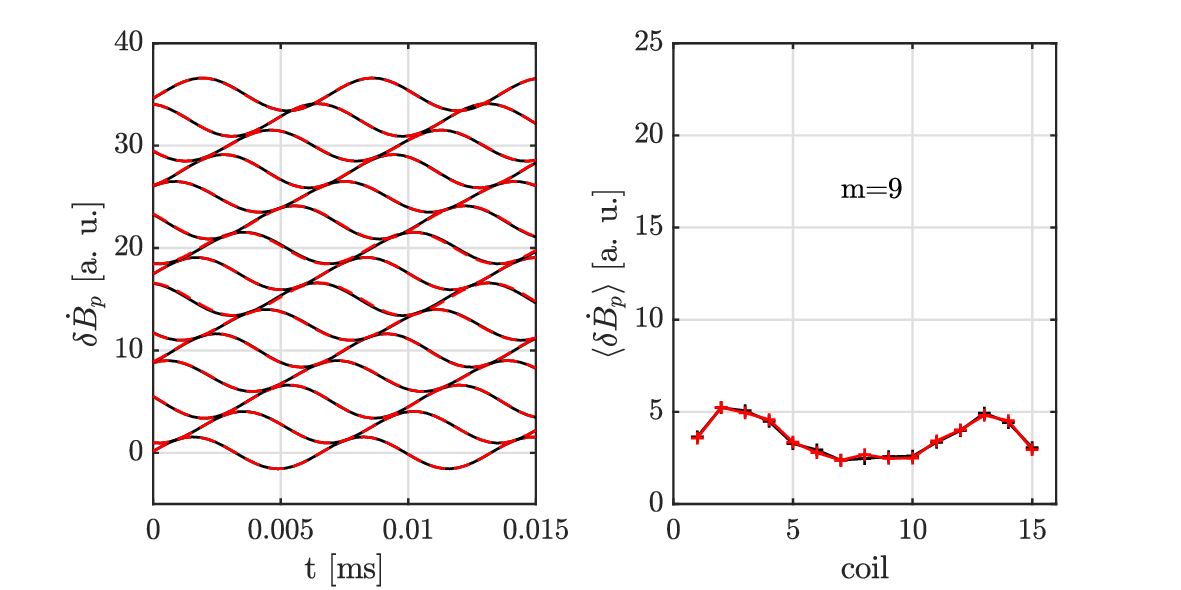}
    \caption{Comparison between the results obtained using the Biot-Savart integral (black lines) and solving the wave equation by Fourier methods (red lines) without plasma response.}
    \label{fig:comp_bs_f}
\end{figure}

\begin{figure}[t]
    \includegraphics[width=\columnwidth]{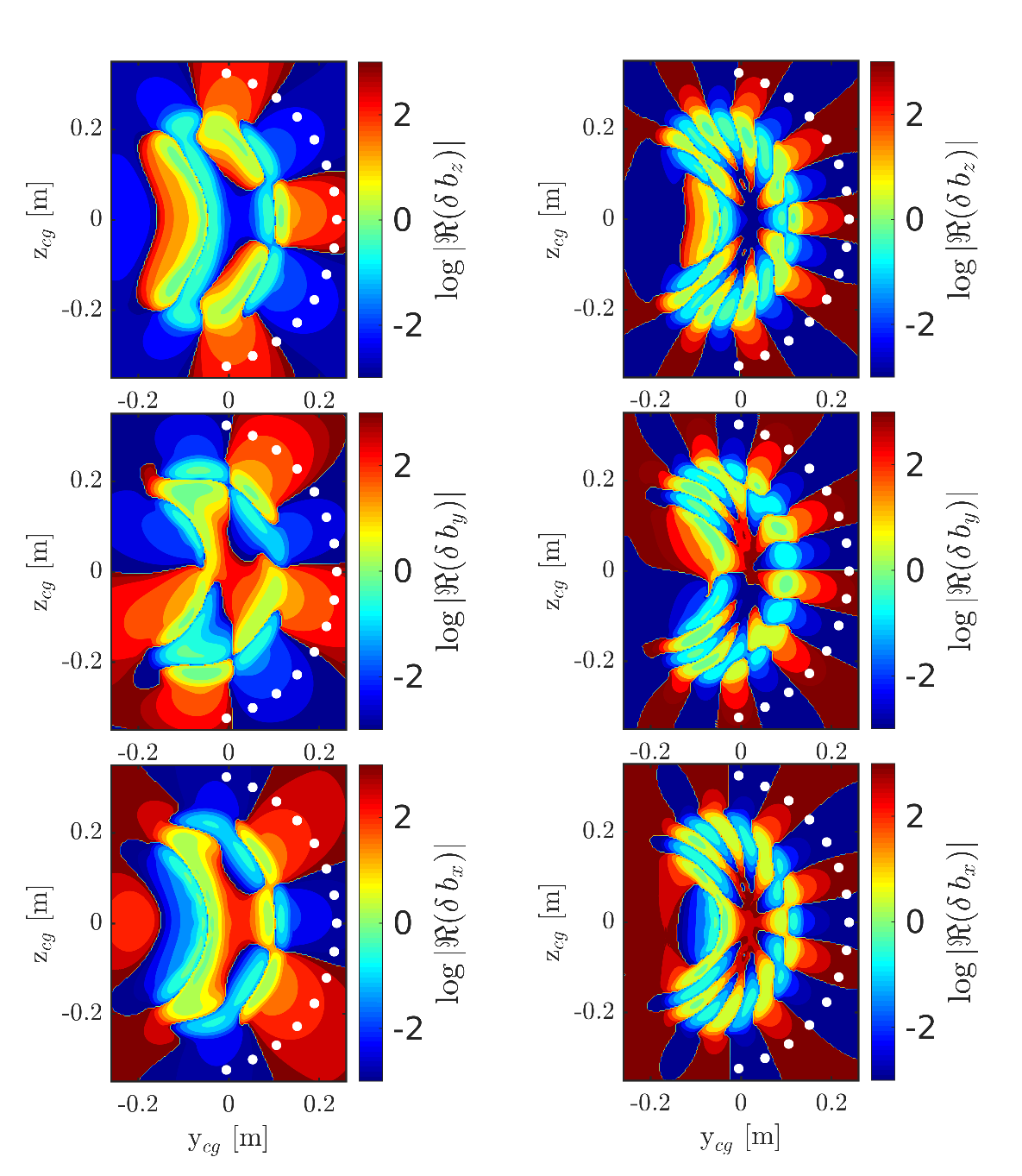}%
    \caption{
        Normalized components of the perturbed magnetic field in the central plane of the simulation grid \mbox{\(x_\text{cg} = 0\)} for \(m=3\) (left) and \(m=9\) (right).
        White circles indicate the location of the poloidal coils.
        By construction, the coils are sentitive only to \(\delta B_z\) and \(\delta B_y\).}
    \label{fig:bfield_NOPR}
\end{figure}

\begin{figure}[h!]
    \includegraphics[width=\columnwidth]{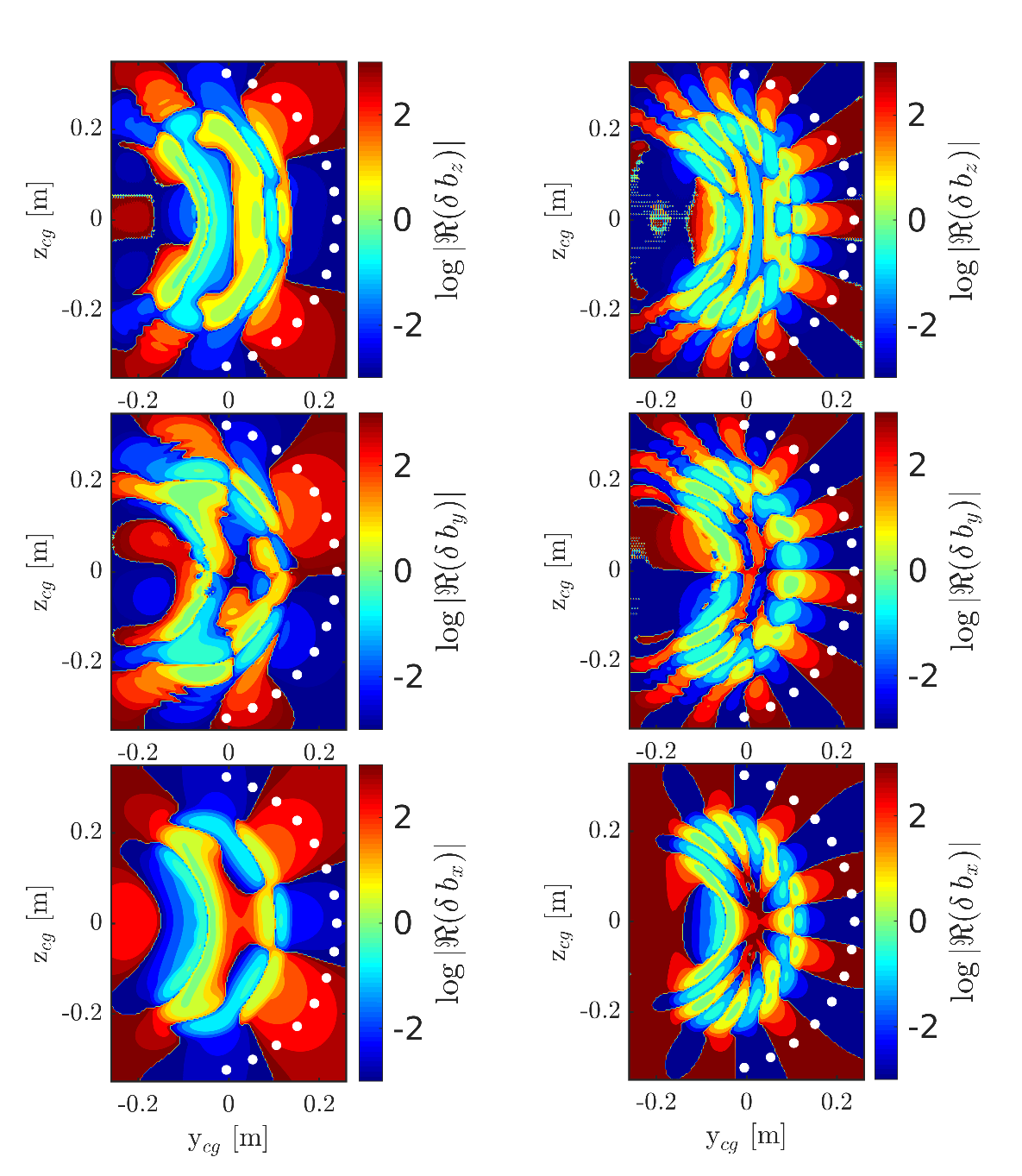}%
    \caption{Same result as in figure \ref{fig:bfield_NOPR} but including now the response of a homogeneous plasma with \(n_e=10^{19} \text{m}^{-3}\) and \(B_0=1\) T.}
    \label{fig:bfield_PR}
\end{figure}

To estimate the effect of the plasma response on the determination of the poloidal mode number, the localized distribution of currents, initially defined in the nodes of the boozer coordinates grid, is interpolated in a cartesian grid.
For convenience, the grid and the current source are then rotated so that the \mbox{\(\vb{x}_\text{cg}\)} direction of the cartesian grid is perpendicular to the plane of the poloidal array, which lies now in the plane formed by vectors \mbox{\(\vb{y}_\text{cg},\vb{z}_\text{cg}\)} of the grid.
The 3D Fourier transform of the current source in this grid, \( \delta\vb{J}_{ext} (\vb{r}) \), is computed numerically using a multidimensional FFT algorithm and used in equation \ref{eq:waveeq_sol}.
Since we want to highlight the effect of the plasma response, the field is only evaluated at those sensors of the poloidal array whose spacelike distribution most closely matches the shape of the plasma (see white dots in figures \ref{fig:bfield_NOPR} and \ref{fig:bfield_PR}).

We first perform a sanity check without plasma response to ensure that the magnetic field calculated at the coil locations using the Biot-Savart integral over the cartesian grid and the solution \ref{eq:field_fourier} based on Fourier decomposition methods yield similar results both in phase and wave amplitude.
Figure \ref{fig:comp_bs_f} shows the normalized synthetic signals obtained with both methods and the wave amplitudes measured by each coil.
Also, figure \ref{fig:bfield_NOPR} shows the real part of the components of the perturbed magnetic field calculated using the Fourier method, for \(n=0, m=3\) and \(n=0, m=9\) modes.
The magnetic field perturbation that arises in response to the mode currents shows a variation of three orders of magnitude between the inner part of the plasma (milliTeslas) and the part where the coils are located (microTeslas) so we have chosen to represent the signed logarithm of the normalized values (\mbox{\(-\text{sign} (\delta b_i) * \log(\abs{\delta b_i})\)}).
This choice of representation emphasizes the structure of the magnetic field perturbation outside the plasma, which is our region of interest.

As discussed in the appendix, solving the equation in Fourier space allows us to describe the plasma response through the dispersion tensor of each plane wave component.
By taking a constant density and a constant static field \(B_0\) along \mbox{\(\vb{x}_\text{cg}\)} in all the space domain we can take into account, approximately, the fast mobility of plasma electrons along the direction of the field and the subsequent shielding that they produce.
Figure \ref{fig:bfield_PR} shows the result of solving equation \ref{eq:waveeq} considering now the contribution of the plasma response to the dispersion tensor given in \ref{eq:dis_tensor}.
A homogeneous plasma with \mbox{\(n_e=10^{19}\ \text{m}^{-3}\)}, immersed in a \mbox{\(B_0=1\)} T field, is taken to explore this effect.

\begin{figure}[h!] \centering
    \includegraphics[]{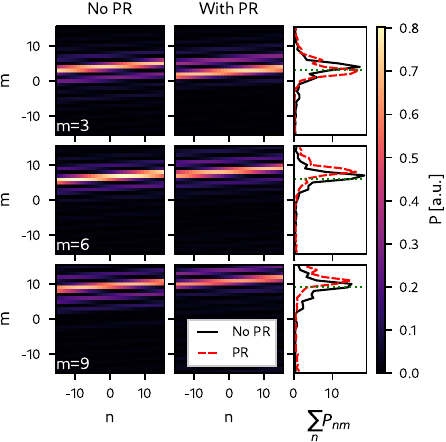}
    \caption{Lomb periodograms obtained with (right) and without plasma response (left) for modes with $n=0$ and $m=3, 6$ and $9$.}
    \label{fig:lombs__3x3__minus_5}
\end{figure}

Evolving in time the field perturbation, for each of the cases represented in figures \ref{fig:bfield_NOPR} and \ref{fig:bfield_PR}, allows us to calculate the synthetic signals and apply the Lomb periodogram analysis.
Figure \ref{fig:lombs__3x3__minus_5} shows the the results of these analysis.
As expected from the result anticipated in figure \ref{fig:comp_bs_f}, the mode structure without plasma response is well diagnosed by the set of coils for all the mode numbers considered.
On the contrary, the plasma response clearly modifies the field structure and consequently the mode number detected by the coils differs from the input ones, the differences \(\Delta m\equiv m_{out}-m_{in}\) reaching values up to \(\Delta m=\pm 2\).
This can be visually observed in figures \ref{fig:bfield_NOPR} and \ref{fig:bfield_PR} where it is shown that the structure of maxima and minima of \(\delta\vb{B}\) is modified by the response of the plasma, in particular for those components of the magnetic field that have an impact on the coils measurement (\(\delta B_z\) and \(\delta B_y\)).
Note that the component of the perturbed field along the direction of the static homogeneous field (\(\delta B_x\)), which is not detected by the coils, is precisely the component less modified by the plasma response since mode currents flow mainly in the \mbox{\(\vb{x}_\text{cg}\)} direction and thus barely contribute to changes in \(\delta B_x\).

The approximations made in the plasma response model (i.e. modes with n=0 and toroidally localized current source), do not allow us to estimate with certainty the changes that the plasma response induces in the measured mode number.
Indeed, the detected mode numbers show differences with respect to the initial ones.
However, this can only serve as an indication that the plasma has a clear effect on mode detection, but we cannot provide a clear quantification of this since the price of assuming an homogeneous plasma is that the electric field associated with the mode produces currents in the whole volume and not only in the part where the plasma is actually present.
This can only be addressed with a full-wave code that solves this problem in real space.

% \FloatBarrier

\section{Phase delay term for the 3D Lomb periodogram}
\label{app:tau_lomb}

The original Lomb periodogram includes a phase delay term in the form of a time shift \(\delta t\) that makes the periodogram independent of time shifts.
This time shift is introduced in \cite{lombLeastsquaresFrequency1976} as a way of cancelling the term:
\begin{equation}
    \mathrm{CS} = \sum_i \cos \omega t_i \sin \omega t_i
\end{equation}
where \(i\) is the sample index.
Changing the model to:
\begin{equation}
    y_i + \epsilon_i = a \cos\omega(t_i - \delta t) + b \sin\omega(t_i - \delta t)
\end{equation}
and solving:
\begin{equation}
    \sum_i \cos \omega (t_i + \delta t) \sin \omega (t_i + \delta t) = 0
\end{equation}
We get an expression for the time shift, that depends on the frequency:
\begin{equation}
    \tan 2\omega\delta t = \dfrac{\sum_i \sin 2\omega t_i}{\sum_i \cos 2\omega t_i}
\end{equation}

For the three-dimensional case, the justification for setting \(\mathrm{CS} = 0\) remains the same.
To achieve that, we discard the time shift idea and instead introduce a phase shift, \(\tau\), that depends on both the mode numbers and the angular frequency.
Changing the model accordingly to:
\begin{equation}
    y_i + \epsilon_i = a \cos (\alpha_{ij} + \tau) + b \sin (\alpha_{ij} + \tau)
\end{equation}
where \(\alpha_{ij}\) is \(\alpha_{ij} \equiv m\theta_j + n\varphi_j - \omega t_i\), and solving again \(\mathrm{CS} = 0\), we get equation \ref{eq:tau_definition}, repeated below:
\begin{equation}
    \tan 2 \tau = \dfrac{\sum_{ij} \sin 2\alpha_{ij}}{\sum_{ij} \cos 2\alpha_{ij}}
\end{equation}
This has the added advantage that the periodogram becomes invariant to both time shifts and angular shifts.

\section{Wave field attenuation} \label{sec:wave_attenuation}

\begin{figure}[h] \centering
    \includegraphics[]{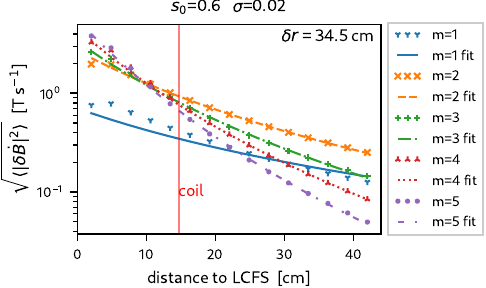}
    \caption{Graph of perturbation intensity (mean squared norm of \(\delta\dot{\vb{B}}\)) vs coil distance to the LCFS for different mode numbers, along with fit to \(1/r^{m+1}\).
    }
    \label{fig:dists}
\end{figure}

We inquire now about the effects of coil distance over the measured signal.
For that, a set of signals with different mode numbers have been simulated for a hypothetical linear array of coils located at \(\varphi=0\), \( z=0\) and distributed with increasing radial distance \(R\).
The amplitude of the perturbed field at the \(i\)-th coil position is
\begin{equation}
    A_i \equiv \sqrt{\Bigl\langle \abs{\delta\dot{\vb{B}_i}}^2 \Bigr\rangle_t} \,,
\end{equation}
where \(\langle\cdot\rangle_t\) denotes a time average.
In the multipolar approximation \cite{jacksonClassicalElectrodynamics1999}, this signal follows an inverse-power law, decaying as \(1/r^{m+1}\), so an attempt has been made to fit the field amplitudes to the function
\begin{equation}
    A(r) = \dfrac{A_0 (m)}{(r_\text{LCFS} + \delta r)^{m+1}} \,,
\end{equation}
where \(A_0 (m)\) is a mode-dependent initial amplitude, \(r_\text{LCFS}\) is the distance between the coil and the LCFS, and \(\delta r\) is a free parameter.
Figure \ref{fig:dists} shows the results of the fit of this function to the simulated perturbation amplitudes for a set of \(n=0\) modes.
For the fit, while \(A_0\) is \(m\)-dependant and thus is fitted separately for each value of \(m\), \(\delta r\) is fitted simultaneously to all signals, ensuring it remains constant.
As the multi-polar approximation is valid only when the measuring point is reasonably far from the plasma, the first coils have been omitted from the fit.
The goodness of the fit for these few first coils improves with increasing mode number, but for \(m=1\) notable deviations can be observed.
The obtained value of \(\delta r\) is consistent with a multipole located approximately over the central field conductor of the device, instead of being over the magnetic axis as expected.
This is probably due to the shape of the plasma column, that differs notably from a cylinder, so some deviation from the ideal case is to be expected.
As a diagnostic, an arrangement like this would be of little use, as the contributions from the toroidal and poloidal modes would be mixed and very difficult to decouple, and would require a space between the LCFS and the first wall that in TJ-II is not available.

\section{Performance}
As the time evolution of the mode can be decoupled from the Biot-Savart integral over the plasma, the code is very performant.
For a \(n_s \times n_\theta \times n_\varphi =  100 \times 150 \times 1000\) grid, the integration of a single mode takes around 5 seconds per coil on an 8-core desktop computer.
The code is mainly written in \texttt{Fortran}, making use of the \texttt{FFTW3} library \cite{frigoDesignImplementation2005} for the calculation of fast Fourier transforms and is parallelized using \texttt{OpenMP}; it also has a \texttt{Python} API for ease of use.

\section*{References}
\bibliography{bib/paper_synth_mirnov.bib}

\end{document}